\documentclass[aps,10pt,prd,twocolumn,superscriptaddress,preprintnumbers,showkeys,nofootinbib]{revtex4-2}
\usepackage{amsmath,amsfonts,amssymb,amscd,amsxtra,amsthm}
\usepackage{graphicx}  
\usepackage{epstopdf} 
\usepackage{dcolumn}  
\usepackage{bm}          
\usepackage{slashed}
\usepackage{cancel}
\usepackage{placeins}
\usepackage{mathtools}
\usepackage{amsbsy}
\usepackage{amstext}
\usepackage{hyperref}

\usepackage{wasysym}
\usepackage[utf8]{inputenc} 
\usepackage{booktabs} 
\usepackage[normalem]{ulem} 
\usepackage[dvipsnames]{xcolor} 
\usepackage{tabularx}
\usepackage{enumitem}  
\usepackage{array} 
\usepackage{slashed}
\usepackage{tikz}
\usepackage{multirow}
\usepackage{cancel}
\usepackage{physics}
\renewcommand\sout{\bgroup\color{red} \ULdepth=-.5ex \ULset}

\makeatletter

\begin{document}  
\preprint{INHA-NTG-02/2026}
\title{Multipole structure of the nucleon tensor form factors}  

\author{Nam-Yong Ghim}
\email[E-mail: ]{Namyong.ghim@inha.edu}
\affiliation{Department of Physics and Institute of Quantum Science,
  Inha University, Incheon 22212, Republic of Korea} 

\author{Ho-Yeon Won}
\email[E-mail: ]{hoyeon.won@polytechnique.edu}
\affiliation{CPHT, CNRS, \'Ecole polytechnique, Institut Polytechnique
  de Paris, Palaiseau, France} 

\author{June-Young Kim}
\email[E-mail: ]{jun-young.kim@inha.ac.kr}
\affiliation{Department of Physics  and Institute of Quantum Science,
  Inha University, Incheon 22212, Republic of Korea}  
\affiliation{CPHT, CNRS, \'Ecole polytechnique, Institut Polytechnique
  de Paris, Palaiseau, France}

\author{Hyun-Chul Kim}
\email[ E-mail: ]{hchkim@inha.ac.kr}
\affiliation{Department of Physics and Institute of Quantum Science,
  Inha University, Incheon 22212, Republic of Korea}
\affiliation{School of Physics, Korea Institute for Advanced Study 
  (KIAS), Seoul 02455, Republic of Korea}
\date{\today}
\begin{abstract}
We investigate the multipole structure of the nucleon tensor form
factors within the chiral quark-soliton model based on the $1/N_c$
expansion. Extending the previous leading-order
analysis~\cite{Ghim:2025gqo}, we include the rotational $1/N_c$
corrections. These corrections provide the leading nonvanishing
contributions to the flavor components that are absent at leading
order, thereby completing the flavor decomposition
of the tensor multipole form factors at the present order.
We numerically evaluate the isoscalar tensor charge,
the isovector anomalous tensor magnetic moment,
and the isoscalar tensor quadrupole moment,
obtaining $g_T^{u+d}=0.81$, $\kappa_T^{u-d}=1.97$,
and $E_T^{u+d}(0)=5.98$, respectively.
The isoscalar tensor charge and quadrupole moment
are mainly governed by the valence-quark contribution,
whereas the isovector anomalous tensor magnetic moment
receives a sizable Dirac-sea contribution.
We also examine the momentum-transfer dependence
of the corresponding form factors.
They decrease monotonically with increasing $-t$.
In particular, the isovector anomalous tensor magnetic form factor 
shows a pronounced falloff in the small-$|t|$ region,
reflecting the importance of the Dirac sea
in the tensor dipole structure.
\end{abstract}

\maketitle
\section{Introduction}
The tensor charges of the nucleon carry several significant physical
implications, which are defined as the forward nucleon matrix elements
of the tensor bilinear quark operators~\cite{Adler:1975he,
  Jaffe:1991kp, Jaffe:1991ra, He:1994gz}. For example, they are
essential for probing new physics 
through precision $\beta$-decay experiments and for setting limits on
the neutron electric dipole moment beyond the Standard
Model~\cite{Erler:2004cx, Pospelov:2005pr, Severijns:2006dr,
  Cirigliano:2013xha, Courtoy:2015haa, Bhattacharya:2015wna,
  Gonzalez-Alonso:2018omy, Antel:2023hkf}. They also play an important
role in the interactions of dark matter with the
nucleon~\cite{Goodman:2010ku, Bishara:2017pfq, Brod:2017bsw,
  Antel:2023hkf, Ma:2024aoc, Glick-Magid:2023uhk,
  Liang:2025kkl}. Originally, the flavor-singlet tensor 
charge was introduced as the first Mellin moment of the transversity
parton distribution function (PDF) $h_{1}(x)$~\cite{Jaffe:1991kp,
  Jaffe:1991ra}, which provides essential information on the spin
structure of the nucleon~\cite{Ralston:1979ys, Kodaira:1979ib,
  Artru:1989zv, Bukhvostov:1984rns, Jaffe:1991kp, Jaffe:1991ra,
  Cortes:1991ja} (see also the reviews~\cite{Barone:2010zz,
  Aidala:2012mv, Diehl:2023nmm}). 

While the leading-twist unpolarized PDF $f_{1}(x)$ and the
longitudinally polarized PDF $g_{1}(x)$ can be measured in 
deep-inelastic scattering~\cite{Whitlow:1991uw, Kodaira:1995be}, the
third leading-twist distribution, the transversely polarized PDF
$h_{1}(x)$, is inaccessible in inclusive deep-inelastic
scattering. Instead, it can be extracted from processes such as the
transverse spin asymmetry $A_{TT}$ in Drell--Yan $p\bar{p}$
reactions~\cite{Efremov:2004qs, PAX:2005leu, Anselmino:2004ki,
  Pasquini:2006iv} and in semi-inclusive deep-inelastic scattering
(SIDIS) at HERMES and COMPASS~\cite{HERMES:2004mhh, COMPASS:2005csq,
  Anselmino:2007fs, Anselmino:2008jk}. More refined extractions were
performed in Refs.~\cite{Anselmino:2013vqa, Kang:2014zza} by combining
SIDIS data with the Collins fragmentation
function~\cite{Collins:1992kk} measured by the Belle
Collaboration~\cite{Belle:2008fdv}. Since then, additional empirical
information on the tensor charges has been
obtained~\cite{Radici:2015mwa, Ye:2016prn, Lin:2017stx,
  Cocuzza:2023oam, Cocuzza:2023vqs, Gao:2025evv}. Moreover,
increasingly precise lattice QCD results have been accumulated over
several decades~\cite{Aoki:1996pi, QCDSF:2006tkx, Alexandrou:2017qyt,
  Gupta:2018lvp, Yamanaka:2018uud, Alexandrou:2019brg,
  Alexandrou:2021oih, Park:2021ypf, Alexandrou:2022dtc,
  Rodekamp:2023wpe, Wang:2025nsd}. 

The chiral-odd generalized parton distributions (GPDs) broaden the
physical meaning of the tensor charges and related form
factors~\cite{Ji:1998pc, Diehl:2001pm, Goeke:2001tz, Diehl:2003ny,
  Belitsky:2005qn, Boffi:2007yc, Kumericki:2016ehc, dHose:2016mda},
since they encode information on both PDFs and form factors. In
particular, the tensor form factors arise as the first Mellin moments of the
chiral-odd GPDs. As discussed in Refs.~\cite{Jaffe:1991kp,
  Jaffe:1991ra}, the flavor-singlet tensor charge counts valence
quarks (quarks minus antiquarks) of opposite transversity. Notably,
sea quarks do not contribute because the tensor bilinear quark
operator is odd under charge conjugation. In contrast, the
axial-vector operator is even under charge conjugation, so the axial
charge receives contributions from sea-quark helicities. At leading
twist, there are four chiral-odd GPDs~\cite{Ji:1998pc,
  Diehl:2001pm}. However, only three independent first Mellin moments
survive, since one of the GPDs vanishes upon integration over the
Bjorken variable $x$. These
three moments correspond to the independent multipole tensor form
factors. In the forward limit, the first becomes the tensor charge
$\delta q$ for quark flavor $q$. The second is the tensor anomalous
magnetic dipole moment $\kappa_{T}^{q}$, which characterizes the
deformation of the average transverse position of quarks polarized in
the transverse direction relative to the transverse center of light-front momentum for an
unpolarized target~\cite{Burkardt:2005hp}.
In
Refs.~\cite{Diehl:2005jf, QCDSF:2006tkx}, the transverse spin
structure of the nucleon was further explored using chiral-odd GPDs,
and the physical interpretation of the final tensor form factor,
$Q_{T}^{q}$, was clarified. These chiral-odd GPDs and tensor form
factors have been studied within various theoretical approaches,
including constituent quark models in the light-cone
basis~\cite{Pasquini:2005dk, Lorce:2011dv}, MIT bag
models~\cite{He:1994gz, Tezgin:2024tfh}, light-front quark
models~\cite{Kaur:2023lun, Kaur:2025ssu}, light-cone quark
models~\cite{Luan:2024vgv}, and QCD sum rules~\cite{Aliev:2011ku}. 

They have also been studied in the chiral-quark soliton model
($\chi$QSM)~\cite{Kim:1995bq, Kim:1996vk, Wakamatsu:2008ki,
  Lorce:2007fa, Ledwig:2010zq, Ledwig:2010tu, Ledwig:2011qw,
  Kim:2024ibz, Ghim:2025gqo, Kim:2025mol}. In our previous
work~\cite{Ghim:2025gqo}, we revisited the tensor form factors,
ensuring consistency between the chiral-odd GPDs~\cite{Kim:2024ibz}
and the tensor form factors in the large-$N_{c}$ limit of QCD. We
elaborated on the anomalous tensor magnetic dipole moments studied in
Refs.~\cite{Ledwig:2010tu, Ledwig:2011qw}, and we computed the leading
contribution to the tensor quadrupole moment $Q_{T}$ in the
large-$N_{c}$ limit. This quantity is as important as the tensor
charge and the anomalous tensor magnetic dipole moment, since it is
also related to the leading-twist chiral-odd GPD $\tilde{H}_{T}$. 

This work extends the investigation of Ref.~\cite{Ghim:2025gqo} by
including rotational $1/N_{c}$ corrections~\cite{Kim:2025mol} within
the framework of the $\chi$QSM. The model has the distinctive
advantage that it satisfies the polynomiality of the chiral-odd GPDs
and various sum rules in the large-$N_{c}$ limit of
QCD~\cite{Kim:2024ibz}. The $\chi$QSM incorporates chiral symmetry and
its spontaneous breaking in QCD, which can be derived from the
instanton vacuum~\cite{Callan:1977gz, Carlitz:1978yj, Shuryak:1981ff,
  Diakonov:1985eg} (see Refs.~\cite{Schafer:1996wv, Diakonov:2002fq}
for reviews). The instanton vacuum is characterized by two parameters,
namely the average instanton size $\bar{\rho}\approx 1/3$ fm and the
mean separation between instantons (and anti-instantons)
$\bar{R}\approx 1$ fm. It generates localized quark zero modes with
definite chirality, which subsequently become delocalized and undergo
chirality flips as quarks propagate through the
instanton--antiinstanton medium. As a result, the quarks acquire a
dynamical mass and the pseudo-Nambu--Goldstone bosons emerge,
realizing the spontaneous breaking of chiral symmetry
(SB$\chi$S). Furthermore, this picture introduces the instanton
packing fraction as a small parameter, enabling a systematic expansion
for studying SB$\chi$S and hadronic correlation functions. Effective
dynamics at the scale $\bar{R}$ can be constructed using $1/N_{c}$
expansion techniques, including the saddle-point approximation and
bosonization. This leads to a description in which quarks with a
dynamical mass $M\sim 0.3$--$0.4$ GeV interact with a chiral pion
field. Within this framework, the nucleon emerges as a bound state of
$N_{c}$ valence quarks in a self-consistent mean
field~\cite{Diakonov:1987ty}. Thus, the $\chi$QSM provides a concrete
realization of the baryon picture in large-$N_{c}$
QCD~\cite{Witten:1979kh, Witten:1983tw}. This effective chiral theory
for the nucleon has been very successful in describing many nucleon  
observables~(see Ref.~\cite{Christov:1995vm} for a review). The model
can also describe low-lying singly heavy baryons on the same
footing~\cite{Yang:2016qdz, Kim:2018cxv, Kim:2019rcx, Suh:2022atr}. 

This work is organized as follows.
We first set up the general formalism of the tensor form factors
in Sec.~\ref{sec:2},
where the multipole expansion is performed in the Breit frame.
After briefly reviewing the basic ingredients of the $\chi$QSM
in Sec.~\ref{sec:3},
we apply the model to the tensor form factors
and derive the relevant expressions in Sec.~\ref{sec:4}.
The numerical results are then presented and discussed
in Sec.~\ref{sec:5}.
Finally, Sec.~\ref{sec:6} summarizes the main findings
and gives the conclusions.

\section{Tensor form factors \label{sec:2}}
The quark tensor operator is defined by 
\begin{align}
O^{\mu \nu}_q(x)
= \bar{\psi}_{q}(x)i\sigma^{\mu\nu}\psi_{q}(x),
\label{eq:tensor_op}
\end{align}
where $\psi_q$ is the quark field operator of flavor $q=u,d,\cdots$,
and $\sigma^{\mu\nu}=i[\gamma^\mu,\gamma^\nu]/2$. We denote the
matrix element of the tensor operator as  
\begin{align}
    \mathcal{M}^{q}[i\sigma^{\mu\nu}]
    =
    \langle     N(p',S'_3)|     O^{\mu \nu}_q(0)    |N(p,S_3)
    \rangle ,
    \label{eq:tensor_me}
\end{align}
where $p$ and $p'$ are respectively the initial and final nucleon four
momenta, and $S_3$ and $S'_3$ are the corresponding spin
projections. The nucleon states are normalized as
\begin{align}
    \langle N(p',S'_3)|N(p,S_3)\rangle
    =
    (2\pi)^3 2p^0\,
    \delta_{S'_3S_3}\,
    \delta^{(3)}(\bm p'-\bm p).
\end{align}
Then Eq.~\eqref{eq:tensor_me} can be parametrized in terms of the
tensor form factors
\begin{align}
    \mathcal{M}^{q}[i\sigma^{\mu\nu}]
    =
    \bar{u}
    \bigg[
    H_T^q\, i\sigma^{\mu\nu}
    &+\tilde{H}_T^q\,
    \frac{P^\mu\Delta^\nu-P^\nu\Delta^\mu}{M_N^2}
    \nonumber\\
    &+E_T^q\,
    \frac{\gamma^\mu\Delta^\nu-\gamma^\nu\Delta^\mu}{2M_N}
    \bigg]
    u,
    \label{eq:tensor_param}
\end{align}
where $u\equiv u(p,S_3)$ and $\bar{u}\equiv \bar{u}(p',S'_3)$. The average
momentum and momentum transfer are defined by
\begin{align}
    P=\frac{p'+p}{2},\qquad \Delta=p'-p .
\end{align}
For on-shell nucleons, $p'^2=p^2=M_N^2$, they obey
\begin{align}
    P\cdot\Delta=0,\qquad
    P^2+\frac{\Delta^2}{4}=M_N^2 ,
        \label{eq:onshell}
\end{align}
where $M_N$ is the nucleon mass. The form factors $F^{q}\equiv
F^{q}(t)$, with $F=H_T,\tilde H_T,E_T$, are functions of $t=\Delta^2$;
their dependence on the renormalization scale is understood and not
displayed explicitly.  

We work in the three-dimensional~(3D) Breit frame~(BF), where the
average momentum and momentum transfer are given by 
\begin{align}
    P=(P^0,\bm{0}), \qquad \Delta=(0,\bm{\Delta}) ,
\end{align}
with $(P^0)^2=M_N^2-t/4$ following from the on-shell
constraint~\eqref{eq:onshell}. In this frame, the matrix
element~\eqref{eq:tensor_param} can be expanded in terms of
irreducible tensors constructed from $\bm{\Delta}$:
\begin{subequations}
\label{eq:3Dmul} 
\begin{align}
    \mathcal{M}^{q}[i\sigma^{0 k}]
& = \bm{1} \,  Y^{k}_{1} \sqrt{-t}
    \left[ 
    H^{q}_{T} 
  + E^{q}_{T} 
  + \frac{2\left(P^{0}\right)^{2}}{M_{N}^{2}}
    \tilde{H}^{q}_{T}
    \right], 
    \label{eq:3Dmul_a} \\ 
    \mathcal{M}^{q}[i\sigma^{ij}] 
& = 2M_{N}i \epsilon^{ijk} \sigma^{k} Y_{0}
    \cr
    &\times \left[ 
    \left(\frac{P^{0}}{3M_{N}}+\frac{2}{3}\right) 
    H^{q}_{T}
  + \frac{t}{6M^{2}_{N}}  
    E^{q}_{T}
    \right] \cr 
&\hspace{-0.5cm} - i \epsilon^{ijk} \sigma^{m} Y^{km}_{2} \frac{t}{2M_{N}} 
    \left[
    \frac{M_{N}}{P^{0}+M_{N}} 
    H^{q}_{T} 
  + E^{q}_{T}
    \right].
\label{eq:3Dmul_b}  
\end{align}
\end{subequations}
Here the rank-$n$ irreducible tensors are defined as
\begin{align}
    Y_0=1,\quad
    Y_1^i=\frac{\Delta^i}{|\bm{\Delta}|},\quad
    Y_2^{ij}
    =
    \frac{\Delta^i\Delta^j}{\bm{\Delta}^2}
    -\frac{1}{3}\delta^{ij},
    \label{eq:IRtensor}
\end{align}
where the spatial indices run over $i,j=1,2,3$, and
\begin{align}
\bm{1}\equiv \delta_{S'_3S_3}, \qquad
\bm{\sigma}\equiv \bm{\sigma}_{S'_3S_3},
\label{eq:pauli}
\end{align}
denote the identity matrix and the Pauli matrices between the initial
and final spin states, respectively.

Equation~\eqref{eq:3Dmul} exhibits three independent tensor multipole 
structures. We define the corresponding 3D tensor multipole moments by
the form-factor combinations at $t=0$~\cite{Ghim:2025gqo}: 
\begin{subequations}\label{eq:3Dmultipolemoments}
\begin{align}
    \label{eq:3Dmultipolemomentsa}
\text{Monopole}    &: \quad H_T^q(0),\\[0.8ex]
\label{eq:3Dmultipolemomentsb}
\text{Dipole}      &: \quad H_T^q(0)+E_T^q(0)+2\tilde{H}_T^q(0),\\
\label{eq:3Dmultipolemomentsc}
\text{Quadrupole}  &: \quad \frac{1}{2}H_T^q(0)+E_T^q(0).
\end{align}
\end{subequations}
For comparison, the corresponding two-dimensional (2D) tensor
multipole moments are defined in Ref.~\cite{Ghim:2025gqo} as
\begin{subequations}\label{eq:2Dmultipolemoments}
\begin{align}
\text{Monopole}    &: \quad H_T^q(0)=g_T^q,\\[0.5ex]
\text{Dipole}      &: \quad
                     E_T^q(0)+2\tilde{H}_T^q(0)=\kappa_T^q,\\[0.5ex] 
\text{Quadrupole}  &: \quad 2\tilde{H}_T^q(0)=Q_T^q .
\end{align}
\end{subequations}

\section{Chiral quark-soliton model \label{sec:3}}
In this section, we briefly review the $\chi$QSM, an effective theory
for baryons, and show how the tensor form factors can be evaluated
within this framework. The low-energy effective QCD partition function
is written in Euclidean space as 
\begin{align}
\mathcal{Z}_{\mathrm{eff}}
&=
\int D\psi D\psi^\dagger DU\,
\exp\left[
\int d^4x\,\psi^\dagger D(U)\psi
\right]
\nonumber\\
&=
\int DU\,\exp\left(-S_{\mathrm{eff}}[U]\right),
\label{eq:partition}
\end{align}
with the single-particle Dirac operator defined as
\begin{align}
    D(U)=i\rlap{/}{\partial}+iMU^{\gamma_5}+i\hat{m}.
    \label{eq:DO}
\end{align}
Integrating out the quark fields in Eq.~\eqref{eq:partition}, we
derive the effective chiral action 
\begin{align}
    S_{\mathrm{eff}}[U]
    =
    -N_c\,\mathrm{Tr}\ln D(U).
    \label{eq:eaction}
\end{align}
Here $M$, $\hat{m}$, and $U$ denote the dynamical quark
mass, the current-quark mass matrix, and the chiral field,
respectively. In the instanton vacuum, the dynamical quark mass is
momentum-dependent, $M(k)$, owing to the finite average instanton size
$\bar\rho$. The instanton form factor suppresses quark modes with  
$k\gtrsim \bar\rho^{-1}$, so that $\bar\rho^{-1}$ sets the natural
ultraviolet scale of the effective theory. In the present work, we use a
constant mass $M(k)=M(0)\equiv M$ and implement this scale through the
proper-time regularization. The cutoff mass is identified
approximately with a scale of order $\bar\rho^{-1}$. The current-quark
mass matrix is given as 
\begin{align}
    \hat{m}=\mathrm{diag}(m_u,m_d),
\end{align}
and isospin symmetry is assumed, $m_u=m_d=m$.

The chiral field $U$ is an SU(2) matrix field containing the pion
fields. It is defined as 
\begin{subequations}
\begin{align}
    U^{\gamma_5}
    &=
    \frac{1+\gamma_5}{2}U
    +
    \frac{1-\gamma_5}{2}U^\dagger,
    \\
    U&=\exp\left(i\pi^a\tau^a\right),
\end{align}
\end{subequations}
where $\pi^a$ denote the pion fields and $\tau^a$ represent the Pauli
matrices in flavor space. To describe a baryon, we use a static pion
field with hedgehog symmetry,
\begin{align}
    U^{\gamma_5}(\bm{x})
    =
    \exp\left[
    i\gamma_5\,{\bm{n}}\cdot\bm{\tau}\,P(r)
    \right],
    \label{eq:Ugamma5}
\end{align}
where $r=|\bm{x}|$ and $n^a=x^a/r$, and the Euclidean $\gamma_5$ matrix is $\gamma_5\equiv\gamma_1\gamma_2\gamma_3\gamma_4$. 
Equivalently, the pion field
has the form $\pi^a(\bm{x})=n^a P(r)$, where $P(r)$ is a radial
profile function satisfying $P(0)=\pi$ and $P(\infty)=0$. In this ansatz,
the isospin direction of the pion field is identified with the radial
direction in coordinate space.

From Eq.~\eqref{eq:DO}, the single-particle Dirac Hamiltonian is
obtained as 
\begin{align}
    H(U)
    =
    \gamma_4\gamma_k\partial_k
    +\gamma_4 M U^{\gamma_5}
    +\gamma_4\hat{m}.
    \label{eq:Hamiltonian}
\end{align}
Because of the hedgehog symmetry, the Hamiltonian is invariant under
combined rotations generated by the total angular momentum $\bm
J=\bm L+\bm S$ and the isospin $\bm T$. The generator of this combined
rotation is the grand spin, 
\begin{align}
    \bm G=\bm J+\bm T,
\end{align}
where $\bm L$ and $\bm S$ are the orbital angular momentum and quark
spin operators. Hence $[H(U),\bm G]=0$, and the single-particle 
eigenstates can be classified by the grand-spin quantum number.

The single-particle eigenfunctions and eigenenergies are determined from
\begin{align}
    &H(U)\Phi_n(\bm{x})=E_n\Phi_n(\bm{x}), \nonumber \\[1ex]
&\text{with} \quad  \Phi_{n}(\bm{x})\equiv \langle \bm{x}| n \rangle,
\label{eq:wf}
\end{align}
where the label $n$ collectively denotes the quantum numbers
\begin{align}
    n=\{G,G_3,\Pi, ... \}.
\end{align}
Here $G$ and $G_3$ stand for the grand spin and its projection,
respectively, $\Pi$ is the parity, and the ellipsis denotes the
remaining radial quantum number. The single-particle energies
associated with these states form the Dirac spectrum in the chiral
background field. This field polarizes the positive- and
negative-energy continua and, for a sufficiently strong field,
produces a discrete bound level  
\begin{align}
|\mathrm{lev}\rangle
\equiv
|n=\{E_{\mathrm{lev}},G=0,G_3=0,\Pi=+\}\rangle
\label{eq:bswf}
\end{align}
with $-M < E_{\mathrm{lev}} < M$. Occupying this bound level with the 
$N_c$ valence quarks with the negative-energy Dirac continuum (or
Dirac sea) yields the ground-state baryon with baryon number
$B=1$. Then the classical energy is obtained by 
\begin{align}
    E_{\mathrm{cl}}[U]
    =
    N_c E_{\mathrm{lev}}[U]
    +
    E_{\mathrm{sea}}[U].
\end{align}
Here the first term comes from the occupied bound level,
whereas $E_{\mathrm{sea}}$ denotes the regularized and
vacuum-subtracted contribution of the filled negative-energy
continuum.  

In the large $N_c$ limit, the functional integral over $U$
is dominated by the configuration that minimizes
$E_{\mathrm{cl}}[U]$. This stationary configuration defines the
self-consistent pion mean field $U_{\rm cl}$,
\begin{equation}
 \left.
 \frac{\delta E_{\rm cl}[U]}{\delta U}
 \right|_{U=U_{\rm cl}}
 =0 .
\end{equation}
Evaluating the energy functional at this field, we obtain the
classical nucleon mass
\begin{equation}
 M_{\rm cl}=E_{\rm cl}[U_{\rm cl}] .
\end{equation}
Details of the self-consistent construction are given in
Ref.~\cite{Christov:1995vm}.

Corrections beyond the leading order in the large $N_c$ expansion
arise from $1/N_c$ quantum fluctuations of the pion field around the 
self-consistent field $U_{\rm cl}$, which are neglected.  
The translational and rotational zero modes must be considered
completely. This will assign proper quantum numbers for the nucleon.
Thus, the collective Hamiltonian is obtained as 
\begin{equation}
 H_{\rm rot}
 =
 \frac{\hat{\bm{S}}^{2}}{2I}
 =
 \frac{\hat{\bm{T}}^{2}}{2I},
 \label{eq:ST}
\end{equation}
where $\hat{\bm{S}}$ and $\hat{\bm{T}}$ are the spin and isospin
operators. The moment of inertia is expressed as 
\begin{equation}
 I =
 \frac{N_c}{6}
 \sum_{\substack{n,{\rm occ}\\ m,{\rm non}}}
 \frac{
 \langle n|\tau^a|m\rangle
 \langle m|\tau^a|n\rangle
 }{E_m-E_n}.
\end{equation}
Here the sum runs over occupied levels $n$ and non-occupied levels $m$
of the Dirac Hamiltonian~\eqref{eq:Hamiltonian}. Since $I=O(N_c)$, the
rotational Hamiltonian is $H_{\rm rot}=O(1/N_c)$ and therefore gives a
subleading correction to the nucleon mass.

Because of the hedgehog symmetry, the collective
Hamiltonian~\eqref{eq:ST} has the form of a spherical top. The baryon 
collective wave functions are therefore its eigenfunctions and can be
written as Wigner $D$-functions carrying the spin and isospin quantum
numbers of the quantized classical nucleon. The explicit representation is given
in Ref.~\cite{Christov:1995vm}. 

\section{Tensor form factors in the chiral quark-soliton
  model \label{sec:4}} 
The nucleon matrix element of the tensor current can be derived from
the three-point correlation function:
\begin{align}
    &\langle N(p',S'_3)|O^{\mu
      \nu}_{q,\mathrm{E}}(0)|N(p,S_3)\rangle\cr 
    &=\frac{1}{Z_{\mathrm{eff}}}\lim_{\mathrm{T\rightarrow
      \infty}}\exp\bigg(-ip_{4}\frac{T}{2}-ip'_{4}\frac{T}{2}\bigg)\cr 
    &\times\int d^{3} \bm{x}d^{3}\bm{y}\exp(-i\bm{p'}\cdot
      \bm{y}+i\bm{p}\cdot \bm{x})\cr 
    &\times\int \mathcal{D}\psi\mathcal{D}\psi^{\dagger}\mathcal{D}U\, 
    J_{N}(\bm{y},T/2)O^{\mu \nu}_{q,\mathrm{E}}(0)J_{N}(\bm{x},-T/2)\cr
    &\times\exp\bigg[\int d^{4}z\, \psi^{\dagger}(z)D(U)\psi(z)\bigg],
    \label{eq:correlationF}
\end{align}
where $O^{\mu\nu}_{q,\mathrm{E}}$ is the Euclidean counterpart of the
tensor operator defined in Eq.~\eqref{eq:tensor_op}. The nucleon
Ioffe-type current $J_N$ consists of the $N_c$ valence quarks 
\begin{align}
    J_{N}(x)=&\frac{1}{N_{c}!}\epsilon_{\alpha_{1}\cdots\alpha_{N_{c}}}
    \Gamma^{\beta_{1}\cdots\beta_{N_{c}}}_{SS_{3}TT_{3}}
    \psi_{\alpha_{1}\beta_{1}}(x)\cdots\cr
    &\times\psi_{\alpha_{N_{c}}\beta_{N_{c}}}(x),
\end{align}
where $\alpha_i$ and $\beta_i$ denote color and spin-flavor
indices, respectively. The matrices
$\Gamma^{\beta_{1}\cdots\beta_{N_{c}}}_{SS_{3}TT_{3}}$ carry quantum
numbers such as spin $S$, isospin $T$, and third 
components $S_3$ and $T_3$ for the nucleon.

We decompose the tensor-current matrix element computed in the
$\chi$QSM into isoscalar and isovector components. In the strict
large $N_c$ limit, its spin--flavor structure selects only the leading
flavor components: the isoscalar part of
$\mathcal{M}^{u+d}[i\sigma^{0k}]$ and the isovector part of
$\mathcal{M}^{u-d}[i\sigma^{ij}]$. The complementary components vanish
at this order; see Ref.~\cite{Ghim:2025gqo} for details. They are
generated once the rotational zero modes are quantized. Although the
corresponding rotational corrections are subleading in the $1/N_c$
expansion, they give the leading nonvanishing contributions to
$\mathcal{M}^{u-d}[i\sigma^{0k}]$ and
$\mathcal{M}^{u+d}[i\sigma^{ij}]$. Thus they complete the flavor 
decomposition of the tensor-current matrix element beyond the strict
large $N_c$ limit. 

After evaluating the Euclidean correlation function in
Eq.~\eqref{eq:correlationF}, we convert the resulting matrix element
from Euclidean to Minkowski conventions before matching it to the
tensor form-factor decomposition in Eq.~\eqref{eq:3Dmul}. The
large $N_c$ scaling of the kinematic variables entering the matrix
element is 
\begin{subequations}
    \label{eq:ki}
\begin{align}
\Delta^{0} &= \mathcal{O}(N^{-1}_{c}), && \bm{\Delta} =
                                          \mathcal{O}(N^{0}_{c}),
  \\[0.5ex] 
P^{0} &= \mathcal{O}(N^{1}_{c}), && \bm{P} = \mathcal{O}(N^{0}_{c}),
\end{align}
\end{subequations}
with $t=\mathcal{O}(N_c^0)$. In our analysis, the nucleon mass is
identified with the classical nucleon mass up to rotational corrections, 
\begin{align}
    M_N = M_{\mathrm{cl}}+\mathcal{O}(N_c^{-1}) .
\end{align}
Keeping this hierarchy, we organize the rotational
contributions in terms of mean-field form factors in the 3D BF 
\begin{subequations}
    \label{eq:tensor_multipole}
\begin{align}
    \mathcal{M}^{u-d}[i \sigma^{0k}]
    &=
    Y_{1}^{k}\,\bm{1}\,\sqrt{-t}\,
    F^{u-d}_{\mathrm{mf},1},
    \\[1ex]
    \mathcal{M}^{u+d}[i \sigma^{ij}]
    &=
    i \epsilon^{ijk} \sigma^{k}\,2M_{\mathrm{cl}}Y_{0}\,
    F_{\mathrm{mf},0}^{u+d}
    \cr
    &\quad
    -i\epsilon^{ijk}\sigma^{m}Y_{2}^{km}
    \frac{t}{2M_{\mathrm{cl}}}F_{\mathrm{mf},2}^{u+d}.
\end{align}
\end{subequations}
Here $F_{\mathrm{mf},0}$, $F_{\mathrm{mf},1}$, and $F_{\mathrm{mf},2}$
denote mean-field form factors, with their dependence on  
$t$ suppressed for brevity. The resulting multipole structures are
consistent with the general decomposition in Eq.~\eqref{eq:3Dmul}. 

The matching must be performed with the large-$N_c$ hierarchy of both
the kinematic factors~\eqref{eq:ki} and the tensor form factors kept
consistently. The tensor form factors scale as~\cite{Ghim:2025gqo} 
\begin{subequations}\label{eq:Ncscaling_FFs}
\begin{align}
&\{H^{u-d}_{T}, E^{u-d}_{T}, \tilde{H}^{u-d}_{T}\}
\sim
\{N^{1}_{c}, N^{3}_{c}, N^{3}_{c} \}, \\[1ex]
&\{H^{u+d}_{T}, E^{u+d}_{T}, \tilde{H}^{u+d}_{T}\}
\sim
\{N^{0}_{c}, N^{2}_{c}, N^{2}_{c} \}.
\end{align}
\end{subequations}
We then relate the mean-field form factors to the tensor form factors by
matching Eq.~\eqref{eq:tensor_multipole} to the general decomposition in
Eq.~\eqref{eq:3Dmul}. This gives
\begin{subequations}\label{eq:ff_relations}
\begin{align}
    H^{u+d}_{T}
    +\frac{t}{6M_{N}^{2}}E^{u+d}_{T}
    &=
    F^{u+d}_{\mathrm{mf},0}, \label{eq:ff_relationsa}
    \\
    H_{T}^{u-d}
    +E_{T}^{u-d}
    +2\tilde{H}_{T}^{u-d}
    &=
    F^{u-d}_{\mathrm{mf},1}, \label{eq:ff_relationsb}
    \\[1ex]
    E_{T}^{u+d}
    &=
    F^{u+d}_{\mathrm{mf},2}. \label{eq:ff_relationsc}
\end{align}
\end{subequations}

With the above relations in mind, we use the mean-field form factors
as the basis for the following analysis. Their explicit expressions
are obtained as 
\begin{subequations}
    \label{eq:mf_ff}
\begin{align}
    F^{u+d}_{\mathrm{mf},0}(t)
    &=
    -\frac{N_{c}}{6I}
    \sum_{\substack{n,\mathrm{non}\\ m,\mathrm{occ}}}
    \frac{\langle n|\tau^{l}|m\rangle}{E_{n}-E_{m}} \cr
    &\times \langle m|
    \Sigma^{l}\gamma^{0}
    j_{0}(|\hat{\bm{x}}|\sqrt{-t})
    |n\rangle ,
    \\[1ex]
    F^{u-d}_{\mathrm{mf},1}(t)
    &=
    -\frac{i M_{\mathrm{cl}}N_{c}}{3I\sqrt{-t}}
    \sum_{\substack{n,\mathrm{non}\\ m,\mathrm{occ}}}
    \frac{\langle n|\tau^{l}|m\rangle}{E_{n}-E_{m}}
    \cr
    &\times
    \langle m|
    j_{1}(|\hat{\bm{x}}|\sqrt{-t})
    Y_{1}^{k}\left(\frac{\hat{\bm{x}}}{|\hat{\bm{x}}|}\right)
    \gamma^{k}\tau^{l}
    |n\rangle ,
    \\[1ex]
    F^{u+d}_{\mathrm{mf},2}(t)
    &=
    -\frac{3M_{\mathrm{cl}}^{2}N_{c}}{It}
    \sum_{\substack{n,\mathrm{non}\\ m,\mathrm{occ}}}
    \frac{\langle n|\tau^{l}|m\rangle}{E_{n}-E_{m}}
    \cr
    &\times \langle m|
    j_{2}(|\hat{\bm{x}}|\sqrt{-t})
    Y_{2}^{lk}\left(\frac{\hat{\bm{x}}}{|\hat{\bm{x}}|}\right)
    \Sigma^{k}\gamma^{0}
    |n\rangle.
\end{align}
\end{subequations}
Here $\Sigma^k=\gamma^0\gamma^5\gamma^k$, and $\hat{\bm{x}}$ denotes 
the position operator. The function $j_L(|\hat{\bm{x}}|\sqrt{-t})$ is
the spherical Bessel function, and the irreducible tensors appearing
in Eq.~\eqref{eq:mf_ff} are defined in position space. The
single-particle sums contain both the discrete-level and Dirac-sea
contributions. The discrete-level contribution is obtained by
restricting the occupied-state sum over $m$ to the bound level
specified in Eq.~\eqref{eq:bswf}. 

To obtain the spatial representation of the mean-field form factors,
we insert the coordinate-space completeness relation into
Eq.~\eqref{eq:mf_ff}. We define $\rho_{T0}^{u+d}(r)$,
$\rho_{T1}^{u-d}(r)$, and $\rho_{T2}^{u+d}(r)$ through 
\begin{subequations}
\label{eq:3D_multopole_formfactor}
\begin{align}
    F^{u+d}_{\mathrm{mf},0}(t)
    &=
    \int d^{3}r\,
    j_{0}\left(r\sqrt{-t}\right)
    \rho_{T0}^{u+d}(r),
    \\
    F^{u-d}_{\mathrm{mf},1}(t)
    &=
    3\int d^{3}r\,
    \frac{j_{1}\left(r\sqrt{-t}\right)}{r\sqrt{-t}}
    \rho_{T1}^{u-d}(r),
    \\
    F^{u+d}_{\mathrm{mf},2}(t)
    &=
    15\int d^{3}r\,
    \frac{j_{2}\left(r\sqrt{-t}\right)}
    {\left(r\sqrt{-t}\right)^{2}}
    \rho_{T2}^{u+d}(r).
\end{align}
\end{subequations}
Their explicit expressions are given in Appendix~\ref{app:a}. These 
distributions are normalized so that their 3D integrals yield
the corresponding tensor multipole moments,
\begin{subequations}
\label{eq:moment_md}
\begin{align}
    F^{u+d}_{\mathrm{mf},0}(0)
    &=
    \int d^{3}r\,\rho_{T0}^{u+d}(r)
    =
    g_{T}^{u+d}, \label{eq:moment_mda}
    \\
    F^{u-d}_{\mathrm{mf},1}(0)
    &=
    \int d^{3}r\,\rho_{T1}^{u-d}(r)
    =
    \kappa_{T}^{u-d}+g_{T}^{u-d}, \label{eq:moment_mdb}
    \\ 
    F^{u+d}_{\mathrm{mf},2}(0)
    &=
    \int d^{3}r\,\rho_{T2}^{u+d}(r)
    =
    E_{T}^{u+d}(0). \label{eq:moment_mc}
\end{align}
\end{subequations}

\section{Numerical results and discussion}
\label{sec:5}
We are now in a position to present the numerical results for the
tensor form factors and their moments, and to discuss them. We first
specify the model parameters used in the self-consistent
calculation. The dynamical quark mass is taken to be
$M=350\,\mathrm{MeV}$, as motivated by the instanton-vacuum picture of
QCD~\cite{Diakonov:2002fq}. We employ the proper-time regularization
scheme, in which the cutoff mass $\Lambda$ and the current quark mass
$m$ are fixed by reproducing the experimental values of the pion decay
constant $f_\pi=93\,\mathrm{MeV}$ and the pion mass
$m_\pi=140\,\mathrm{MeV}$, respectively. This yields 
\begin{align}
    \Lambda=643\,\mathrm{MeV}, \qquad m=16\,\mathrm{MeV}.
    \label{eq:parameter}
\end{align}
Further details of the parameter fixing are given in
Ref.~\cite{Christov:1995vm}. 

With these parameters, we solve the Dirac eigenvalue problem in
Eq.~\eqref{eq:Hamiltonian} using the Kahana--Ripka
method~\cite{Kahana:1984be}. The classical energy functional
constructed from the resulting spectrum is minimized to determine the
self-consistent pion mean field. Evaluating the energy at this minimum
yields the classical nucleon mass,
\begin{align}
    M_\mathrm{cl}=1254\,\mathrm{MeV}.
\end{align}
The same finite-box basis is used to evaluate the tensor form
factors. In our previous work~\cite{Ghim:2025gqo}, a spherical box of
radius $D=11\,\mathrm{fm}$ was sufficient. In the present calculation,
however, $E_T^{u+d}(0)$ is sensitive to the large-distance region. We
therefore increase the box size to $D=56\,\mathrm{fm}$ to ensure
numerical convergence. The resulting tensor moments are listed in
Table~\ref{Numerical data} and are compared with those from other
models in Table~\ref{table:2}.
\begin{table}[htp]
\begin{tabular}{ c |c c c c }
\hline
\hline
 & Level & Dirac Sea & Total \\
\hline
$g_{T}^{u+d}$     & 0.78 & 0.03 & 0.81 \\
$\kappa_{T}^{u-d}$& 1.56 & 0.41 & 1.97 \\
$E_{T}^{u+d}(0)$    & 5.66 & 0.32 & 5.98 \\
\hline
\hline
\end{tabular}
\caption{Numerical results for the isoscalar tensor charge $g_{T}^{u+d}$, the
isovector tensor anomalous magnetic moment $\kappa_{T}^{u-d}$, and the
isoscalar 3D tensor quadrupole moment $E_{T}^{u+d}(0)$ of the nucleon. The
results are obtained with $M=350\,\mathrm{MeV}$ and
$m_{\pi}=140\,\mathrm{MeV}$.}
\label{Numerical data}
\end{table}
As shown in Table~\ref{Numerical data}, the discrete level
provides the dominant contribution to the isoscalar tensor charge,
$g_T^{u+d}[\mathrm{lev}]=0.78$, whereas the Dirac sea contributes
only marginally, $g_T^{u+d}[\mathrm{sea}]=0.03$. The latter amounts to
only about $2$--$3\%$ of the total value, $g_T^{u+d}=0.81$. This
result is close to the MIT bag-model estimates,
$g_T^{u+d}=0.88$~\cite{He:1994gz},
$g_T^{u+d}=0.80$~\cite{WAKAMATSU2007398}, and
$g_T^{u+d}=0.82$~\cite{Tezgin:2024tfh}. Since the bag model is based
mainly on valence-quark degrees of freedom, this agreement supports
the dominance of the valence contribution to the isoscalar tensor
charge. Lattice QCD calculations give smaller values, 
$g_T^{u+d}=0.563\pm0.026$~\cite{Alexandrou:2019brg} and
$g_T^{u+d}=0.541\pm0.067$~\cite{Bhattacharya:2015wna}. The present
result is larger than these values by about $30\%$. In the current 
calculation, however, the isoscalar tensor charge is evaluated at the
leading nonvanishing order in the $1/N_c$ expansion. The discrepancy
may therefore be attributed, at least in part, to subleading $1/N_c$
corrections. 

For the isovector tensor anomalous magnetic moment, the discrete-level
contribution is $\kappa_T^{u-d}[\mathrm{lev}]=1.56$, while the
Dirac-sea contribution is $\kappa_T^{u-d}[\mathrm{sea}]=0.41$. The
latter accounts for about $21\%$ of the total value and is therefore
sizable compared with the sea contributions to the other tensor
moments. The total value is $\kappa_T^{u-d}=1.97$, which is
larger than the light-front constituent quark model estimates.
Using a harmonic-oscillator wave function, Ref.~\cite{Pasquini:2005dk}
obtained $\kappa_T^{u-d}=1.24$, while the hypercentral potential gives
$\kappa_T^{u-d}=0.81$. This difference is not unexpected, since constituent
quark models are based primarily on valence degrees of freedom, whereas the
present calculation contains a sizable contribution from the polarized Dirac
sea. Lattice QCD gives $\kappa_T^{u-d} = 1.051 \pm
0.094$~\cite{Alexandrou:2022dtc}, which is substantially smaller than
our result. The difference may reflect several systematic effects,
including the model scale of the $\chi$QSM and the truncation of the
$1/N_c$ expansion. In addition, a direct comparison with lattice QCD
requires the tensor operator to be matched at the same renormalization
scale. We therefore regard this comparison as qualitative. Finally,
the light-cone QCD sum-rule calculation of 
Ref.~\cite{Erkol:2011iw} gives $\kappa_T^{u-d}=1.82$, which is in close
agreement with our total value.

The isoscalar tensor quadrupole moment receives a discrete-level
contribution of $E_T^{u+d}(0)[\mathrm{lev}]=5.66$ and a Dirac-sea
contribution of $E_T^{u+d}(0)[\mathrm{sea}]=0.32$. The sea
contribution is therefore about $5.5\%$ of the discrete-level
contribution, showing that $E_T^{u+d}(0)$ is dominated by the valence
level. The total value is $E_T^{u+d}(0)=5.98$. This behavior differs
from that of the isovector tensor quadrupole moment, for which the
Dirac-sea contribution amounts to almost half of the total
value~\cite{Ghim:2025gqo}. A recent bag-model
calculation~\cite{Tezgin:2024tfh} also finds a positive value for the
tensor form factor $E_T(0)$. Although its magnitude is smaller than
our result, the sign of the quadrupole moment is consistent with the
present calculation. 

\begin{figure}[htp]
    \includegraphics[scale=0.24]{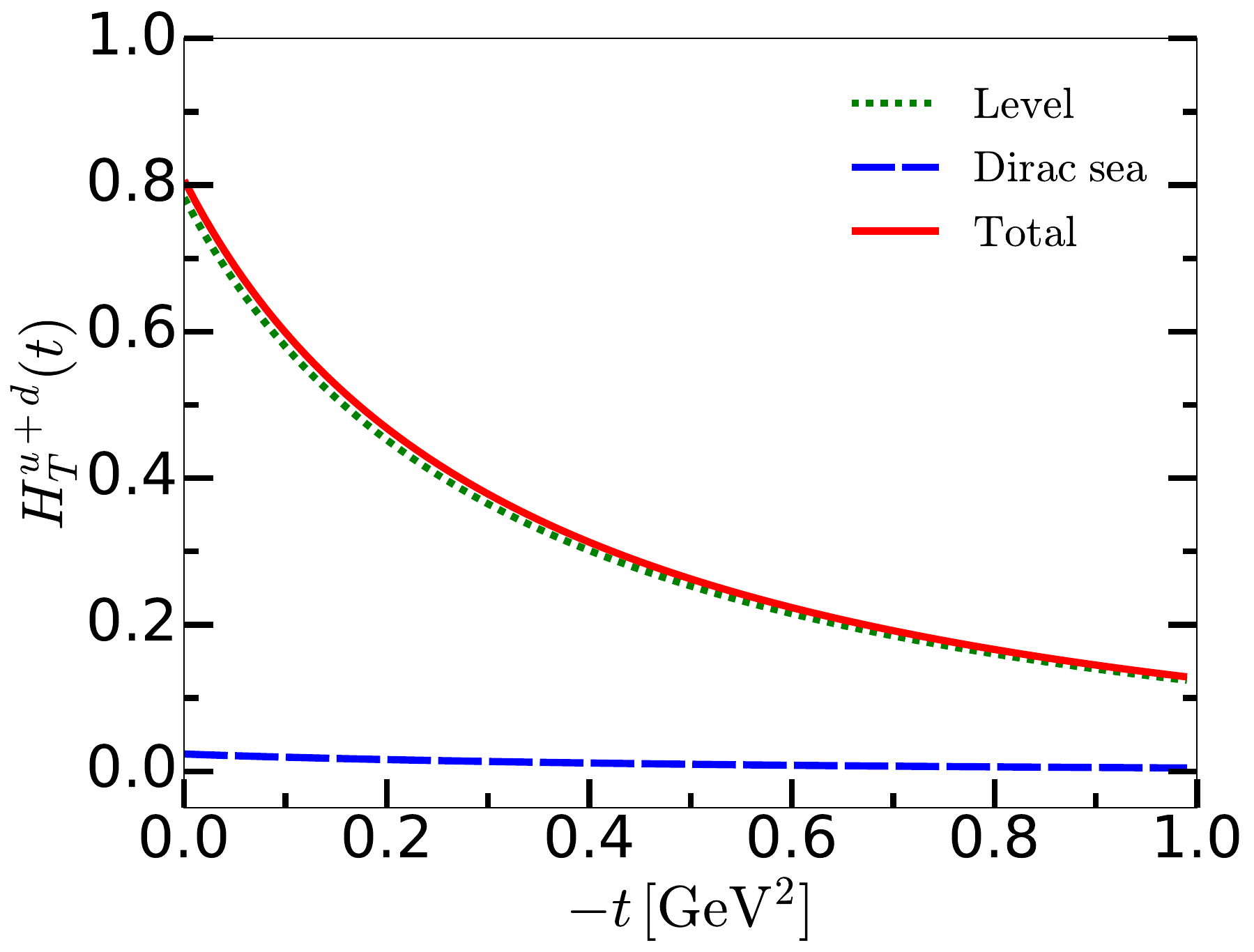}
    \includegraphics[scale=0.24]{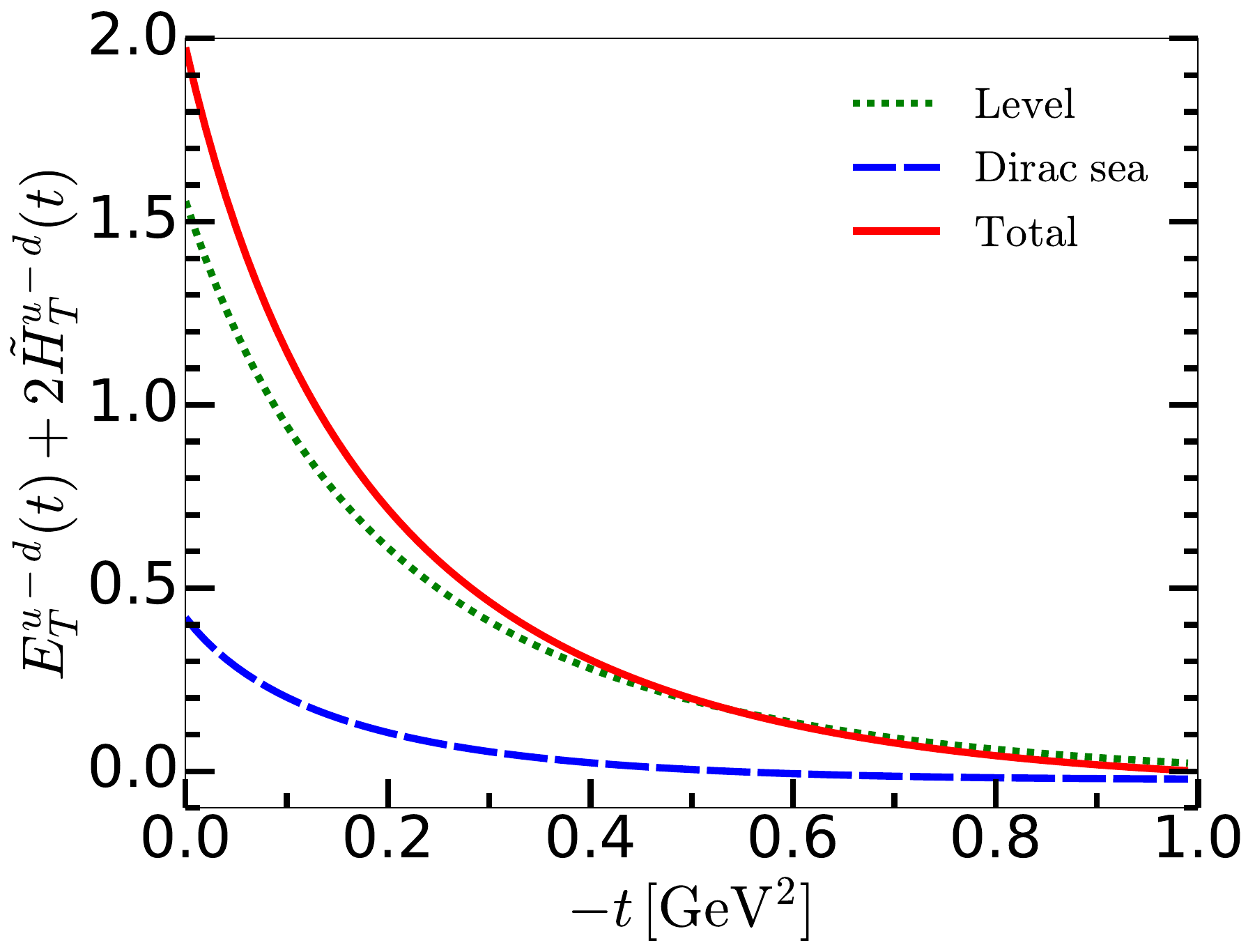}
    \includegraphics[scale=0.24]{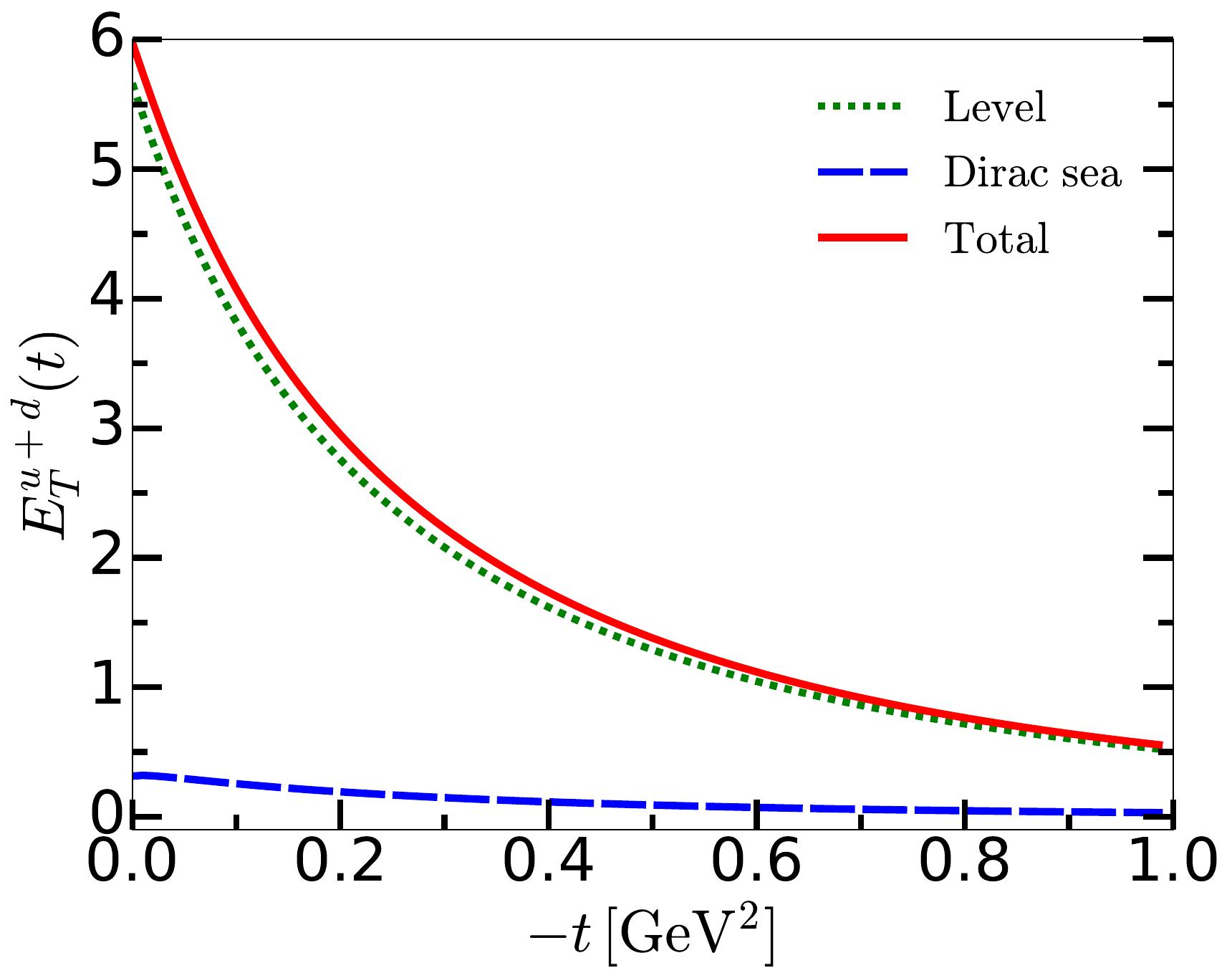}
    \caption{The $t$ dependence of the tensor form factors
$(H_T^{u+d},\,E_T^{u-d}+2\tilde H_T^{u-d},\,E_T^{u+d})$.
Solid curves: total contributions, including the discrete-level and Dirac-sea
parts. Dashed curves: Dirac-sea contributions. Dotted curves: discrete-level
contributions.}  
      \label{fig:1}
\end{figure}

\begin{table}[htp]
\begin{tabular}{ c |c c c}
\hline
\hline
 & $g_{T}^{u+d}$ & $\kappa_{T}^{u-d}$ & $E_{T}^{u+d}$ \\
\hline
 This work & 0.81 & 1.97 & 5.98 \\
 Lattice QCD~\cite{Alexandrou:2019brg} & $0.563\pm 0.026$ &
  $\dots$ & $\dots$ \\
 Lattice QCD~\cite{Bhattacharya:2015wna}  & $0.541\pm 0.067$ &
  $\dots$ & $\dots$ \\
 Lattice QCD~\cite{Alexandrou:2022dtc}  & $\dots $ &
  $1.051\pm 0.094$ & $\dots$ \\
 Bag model~\cite{WAKAMATSU2007398} & $0.80$ &
  $\dots$ & $\dots$ \\
 Bag model~\cite{He:1994gz} & $0.88$ &
  $\dots$ & $\dots$ \\
 Bag model~\cite{Tezgin:2024tfh} & $0.82\pm 0.12$ &
  $\dots$ & $3.7$ \\
 CQM~\cite{Pasquini:2005dk} (HO) & $\dots$ 
  & $1.24$ & $\dots$ \\   
 CQM~\cite{Pasquini:2005dk} (HYP) & $\dots$ 
  & $0.81$ & $\dots$ \\   
 Sum rule~\cite{Erkol:2011iw} & $\dots$ 
 & 1.82 & $\dots$ \\
\hline
\hline
\end{tabular}
\caption{The mean-field results for the tensor form factors compared
  with those from lattice QCD~\cite{Alexandrou:2019brg,
    Bhattacharya:2015wna, Alexandrou:2022dtc},  the bag
model~\cite{WAKAMATSU2007398,He:1994gz,Tezgin:2024tfh}, 
the light-front constituent quark model (CQM)~\cite{Pasquini:2005dk},
and QCD sum rules~\cite{Erkol:2011iw}. For the lattice calculations,
Alexandrou \textit{et al.}~\cite{Alexandrou:2019brg,Alexandrou:2022dtc}
used the physical pion mass, whereas
Bhattacharya \textit{et al.}~\cite{Bhattacharya:2015wna} used
ensembles with $m_\pi=130$, $220$, and $310~\mathrm{MeV}$.}   
\label{table:2}
\end{table}
The tensor multipole moments are given by the forward limits of the
tensor form factors, while their $t$ dependence provides additional
information on the nucleon structure. We therefore evaluate the tensor
form factors using Eq.~\eqref{eq:mf_ff}. Since the large $N_c$ 
expansion is reliable only for $|t|\ll M_N^2$, we restrict ourselves to
$|t|\lesssim 1\,\mathrm{GeV}^2$. The numerical results are displayed in
Fig.~\ref{fig:1}, where the solid curves show the total results, while the
dotted and dashed ones correspond to the discrete-level and Dirac-sea
contributions, respectively.

The Dirac-sea contributions to $H_T^{u+d}$ and $E_T^{u+d}$ decrease
monotonically with increasing $-t$ and remain much smaller than the
corresponding discrete-level contributions. These form factors are
therefore largely governed by the valence level. By contrast, the
Dirac-sea contribution to $E_T^{u-d}+2\tilde{H}_T^{u-d}$ exhibits a
pronounced falloff already in the small-$|t|$ region and is sizable
compared with the sea contributions to $H_T^{u+d}$ and
$E_T^{u+d}$. This indicates that the polarized Dirac sea plays a
non-negligible role in the dipole structure.

\section{Summary and conclusions}
\label{sec:6}
In this work, we have investigated the multipole structure
of the nucleon tensor form factors
within the framework of the chiral quark-soliton model.
This work extends our previous leading-order study
in the $1/N_c$ expansion,
in which the dominant tensor multipole form factors were obtained.
In the leading-order treatment,
the classical nucleon is described by a static pion mean field.
Owing to the spin-flavor symmetry of this mean field,
only selected isospin combinations contribute
to the tensor multipole form factors.
Thus, while the leading-order result captures
the dominant structures,
it does not complete the flavor decomposition.
To obtain the missing flavor components,
one has to include the collective rotational motion
of the classical nucleon.
These rotations generate the rotational $1/N_c$ corrections,
which provide the first nonvanishing contributions
to the isospin components that are absent at leading order.
In the present work,
we computed these contributions explicitly
and thereby obtained the subleading tensor multipole form factors.
We then examined the numerical size of these contributions
through the tensor multipole moments.
The resulting values are
$g_T^{u+d} = 0.81$,
$\kappa_T^{u-d} = 1.97$,
and $E_T^{u+d}(0) = 5.98$.
By separating the discrete-level and Dirac-sea contributions,
we found that the isoscalar tensor charge $g_T^{u+d}$
and the isoscalar tensor quadrupole moment $E_T^{u+d}(0)$
are largely dominated by the discrete level.
This explains why these observables are close
to the quark-model picture based mainly on valence-quark degrees of
freedom.
The isovector anomalous tensor magnetic moment $\kappa_T^{u-d}$,
however, exhibits a different behavior.
It receives a sizable Dirac-sea contribution,
indicating that sea-quark effects are more relevant
for the isovector tensor dipole structure
than for the isoscalar monopole and quadrupole moments.
This pattern is also reflected
in the momentum-transfer dependence
of the corresponding tensor form factors.
In the region considered here,
all three form factors decrease monotonically
as $-t$ increases.
Among them,
the isovector anomalous tensor magnetic form factor,
$E_T^{u-d}(t)+2\tilde H_T^{u-d}(t)$,
exhibits a particularly rapid falloff
in the small-$|t|$ region.
This behavior is consistent with its sizable Dirac-sea contribution
and further supports the importance of sea-quark effects
in the isovector tensor dipole structure.

As a future direction, the present framework can be
extended in several ways. First, the rotational $1/N_c$
expansion also induces corrections to the leading tensor
form factors themselves. Although these contributions are
subleading, they are needed to complete the flavor
decomposition of the tensor multipole form factors in a
consistent manner at the same order in the $1/N_c$
expansion. Their numerical evaluation, however, requires a
careful treatment of the spectral sums over the quark
single-particle states. In particular, the rotational $1/N_c$
correction to the isovector tensor quadrupole form factor is
highly sensitive to finite-box effects in the Kahana--Ripka
basis, which makes its numerical convergence difficult to
control. We therefore leave a systematic calculation of this
correction for future work.
Second, the present framework can be generalized to flavor
SU(3) symmetry. Such an extension would allow us to include
the strange-quark contribution and thereby complete the
flavor decomposition beyond the two-flavor sector. It would
also make it possible to investigate the tensor form factors
of the lowest-lying hyperons within the same theoretical
framework.
Finally, the $\Delta$ baryon appears as a rotationally
excited state of the classical nucleon and can be treated on the same
footing as the nucleon. The study of the $\Delta$ tensor
form factors would therefore provide access to a richer
tensor structure and clarify how the tensor multipole
patterns change under rotational excitation.

\section*{Acknowledgments}
H.-Ch. K. is grateful to C\'{e}dric Lorc\'{e} for valuable discussion
and hospitality during his visit to Le Centre de Physique
Th\'{e}orique (CPHT) at \'{E}cole polytechnique, where part of
the present work was done. The work was supported by the Basic Science
Research Program through the National Research Foundation of Korea
funded by the Korean government (Ministry of Education, Science 
and Technology, MEST), Grants No. RS-2025-00513982 (H.-Ch.K.) and
Grant No. RS-2025-25430144 (N.-Y.Gh.).
H.-Y. W. acknowledges France Excellence scholarship through
Campus France funded by the French government (Ministère de l’Europe
et des Affaires Étrangères), Grant No. 141295X.

\appendix
\section{3D tensor multipole distributions \label{app:a}}
In this appendix, we give the explicit expressions for the spatial
distributions introduced in Eq.~\eqref{eq:3D_multopole_formfactor} in terms
of irreducible matrix elements:
\begin{widetext}
\begin{subequations}
\begin{align}
    \rho_{T0}^{u+d}(r)&=
    -\frac{N_{c}}{6I}
    \bigg(\sum_{E_{n}\neq E_{\mathrm{lev}} }
    \frac{\langle \mathrm{lev}||\tau_{1}|| n\rangle}{E_{\mathrm{lev}}-E_{n}}
    \langle n||r\rangle\{O_{T0} \}_{1} \langle r|| \mathrm{lev} \rangle
    +\sum_{n,m}R^{nm}_{3}
    \langle n||\tau_{1}|| m\rangle
    \langle m||r\rangle\{O_{T0} \}_{1} \langle r|| n \rangle\bigg),
    \\
    \rho_{T1}^{u-d}(r)&=
    -\frac{i M_{\mathrm{cl}}N_{c}}{3\sqrt{3}I}
    \bigg(\sum_{E_{n}\neq E_{\mathrm{lev}}}
    \frac{\langle \mathrm{lev}||\tau_{1}|| n\rangle}{E_{\mathrm{lev}}-E_{n}}
    \langle n||r\rangle \{O_{T1} \}_{1} \langle r|| \mathrm{lev} \rangle
    +\sum_{n,m}R^{nm}_{3}
    \langle n||\tau_{1}|| m\rangle
    \langle m||r\rangle \{O_{T1} \}_{1} \langle r|| n \rangle\bigg),
    \\
    \rho_{T2}^{u+d}(r)&=
    -\frac{2M_{\mathrm{cl}}^{2}N_{c}}{15I}\bigg(
    \sum_{E_{n}\neq E_{\mathrm{lev}}}
    \frac{\langle \mathrm{lev}||\tau_{1}|| n\rangle}{E_{\mathrm{lev}}-E_{n}}
    \langle n||r\rangle \{O_{T2} \}_{1} \langle r|| \mathrm{lev} \rangle
    +\sum_{n,m}R^{nm}_{3}
    \langle n||\tau_{1}|| m\rangle
    \langle m||r\rangle \{O_{T2} \}_{1} \langle r|| n \rangle\bigg).
\end{align}
\end{subequations}
\end{widetext}
Here, the first term in each expression is the discrete-level contribution,
while the second term is the Dirac-sea contribution.
The Dirac-sea contribution
is regularized by the function $R^{nm}_{3}$ in the proper-time
regularization scheme, with the cutoff mass $\Lambda$ given in
Eq.~\eqref{eq:parameter}. The explicit form of $R^{nm}_{3}$ is
\begin{align}
    &R^{nm}_{3}=
    \frac{1}{2\sqrt{\pi}}\int^{\infty}_{1/\Lambda^{2}}
    \frac{du}{\sqrt{u}}\cr
    &\times\bigg[
     \frac{e^{-uE^{2}_{n}}-e^{-uE_{m}^{2}}}{u(E_{m}^{2}-E_{n}^{2})}
    -\frac{E_{n}e^{-uE_{n}^{2}}+E_{m}e^{-uE_{m}^{2}}}{E_{m}-E_{n}} \bigg].
\end{align}
The operators $\{O_{T0}\}_{1}$, $\{O_{T1}\}_{1}$, and $\{O_{T2}\}_{1}$
appearing in the spatial distributions have the following explicit forms:
\begin{subequations}
\begin{align}
  \{O_{T0} \}_{1}&=\Sigma_{1} \gamma^{0}, \\[1ex]
  \{O_{T1} \}_{1}&=[[4\pi rY_{1}\otimes\gamma_{1}]_{0}\otimes\tau]_{1},\\[0.5ex]
  \{O_{T2} \}_{1}&= \left[4\pi r^{2} Y_{2}\otimes\Sigma_{1}\right]_{1}\gamma_{0}.
\end{align}
\end{subequations}

\bibliography{TFF.bib}

@article{Jaffe:1991kp,
    author = "Jaffe, R. L. and Ji, Xiang-Dong",
    title = "{Chiral odd parton distributions and polarized Drell-Yan}",
    reportNumber = "MIT-CTP-1952",
    doi = "10.1103/PhysRevLett.67.552",
    journal = "Phys. Rev. Lett.",
    volume = "67",
    pages = "552--555",
    year = "1991"
}

@article{Whitlow:1991uw,
    author = "Whitlow, L. W. and Riordan, E. M. and Dasu, S. and Rock, Stephen and Bodek, A.",
    title = "{Precise measurements of the proton and deuteron structure functions from a global analysis of the SLAC deep inelastic electron scattering cross-sections}",
    reportNumber = "SLAC-PUB-5442",
    doi = "10.1016/0370-2693(92)90672-Q",
    journal = "Phys. Lett. B",
    volume = "282",
    pages = "475--482",
    year = "1992"
}

@article{Kodaira:1979ib,
    author = "Kodaira, Jiro and Matsuda, Satoshi and Sasaki, Ken and Uematsu, T.",
    title = "{QCD Higher Order Effects in Spin Dependent Deep Inelastic Electroproduction}",
    reportNumber = "RIFP-360",
    doi = "10.1016/0550-3213(79)90329-8",
    journal = "Nucl. Phys. B",
    volume = "159",
    pages = "99--124",
    year = "1979"
}

@article{Kodaira:1995be,
    author = "Kodaira, Jiro",
    editor = "Kodaira, J. and Matsui, T. and Oka, M. and Suzuki, T.",
    title = "{Perturbative QCD and nucleon structure functions}",
    eprint = "hep-ph/9501381",
    archivePrefix = "arXiv",
    reportNumber = "HUPD-9504",
    doi = "10.1143/PTPS.120.37",
    journal = "Prog. Theor. Phys. Suppl.",
    volume = "120",
    pages = "37--56",
    year = "1995"
}

@article{HERMES:2004mhh,
    author = "Airapetian, A. and others",
    collaboration = "HERMES",
    title = "{Single-spin asymmetries in semi-inclusive deep-inelastic scattering on a transversely polarized hydrogen target}",
    eprint = "hep-ex/0408013",
    archivePrefix = "arXiv",
    reportNumber = "DESY-04-141",
    doi = "10.1103/PhysRevLett.94.012002",
    journal = "Phys. Rev. Lett.",
    volume = "94",
    pages = "012002",
    year = "2005"
}

@article{Diakonov:1985eg,
    author = "Diakonov, Dmitri and Petrov, V. Yu.",
    title = "{A Theory of Light Quarks in the Instanton Vacuum}",
    reportNumber = "LENINGRAD-85-1053",
    doi = "10.1016/0550-3213(86)90011-8",
    journal = "Nucl. Phys. B",
    volume = "272",
    pages = "457--489",
    year = "1986"
}

@article{COMPASS:2005csq,
    author = "Alexakhin, V. Yu. and others",
    collaboration = "COMPASS",
    title = "{First measurement of the transverse spin asymmetries of the deuteron in semi-inclusive deep inelastic scattering}",
    eprint = "hep-ex/0503002",
    archivePrefix = "arXiv",
    reportNumber = "CERN-PH-EP-2005-003, DAPNIA-05-17",
    doi = "10.1103/PhysRevLett.94.202002",
    journal = "Phys. Rev. Lett.",
    volume = "94",
    pages = "202002",
    year = "2005"
}

@article{Ji:1998pc,
    author = "Ji, Xiang-Dong",
    title = "{Off forward parton distributions}",
    eprint = "hep-ph/9807358",
    archivePrefix = "arXiv",
    reportNumber = "UMD-PP-98-092, DOE-ER-40762-144",
    doi = "10.1088/0954-3899/24/7/002",
    journal = "J. Phys. G",
    volume = "24",
    pages = "1181--1205",
    year = "1998"
}

@article{Diehl:2001pm,
    author = "Diehl, M.",
    title = "{Generalized parton distributions with helicity flip}",
    eprint = "hep-ph/0101335",
    archivePrefix = "arXiv",
    reportNumber = "DESY-01-009",
    doi = "10.1007/s100520100635",
    journal = "Eur. Phys. J. C",
    volume = "19",
    pages = "485--492",
    year = "2001"
}

@article{Diehl:2003ny,
    author = "Diehl, M.",
    title = "{Generalized parton distributions}",
    eprint = "hep-ph/0307382",
    archivePrefix = "arXiv",
    reportNumber = "DESY-THESIS-2003-018",
    doi = "10.1016/j.physrep.2003.08.002",
    journal = "Phys. Rept.",
    volume = "388",
    pages = "41--277",
    year = "2003"
}

@article{Goeke:2001tz,
    author = "Goeke, K. and Polyakov, Maxim V. and Vanderhaeghen, M.",
    title = "{Hard exclusive reactions and the structure of hadrons}",
    eprint = "hep-ph/0106012",
    archivePrefix = "arXiv",
    doi = "10.1016/S0146-6410(01)00158-2",
    journal = "Prog. Part. Nucl. Phys.",
    volume = "47",
    pages = "401--515",
    year = "2001"
}

@article{Belitsky:2005qn,
    author = "Belitsky, A. V. and Radyushkin, A. V.",
    title = "{Unraveling hadron structure with generalized parton distributions}",
    eprint = "hep-ph/0504030",
    archivePrefix = "arXiv",
    reportNumber = "JLAB-THY-04-34",
    doi = "10.1016/j.physrep.2005.06.002",
    journal = "Phys. Rept.",
    volume = "418",
    pages = "1--387",
    year = "2005"
}

@article{Boffi:2007yc,
    author = "Boffi, Sigfrido and Pasquini, Barbara",
    title = "{Generalized parton distributions and the structure of the nucleon}",
    eprint = "0711.2625",
    archivePrefix = "arXiv",
    primaryClass = "hep-ph",
    doi = "10.1393/ncr/i2007-10025-7",
    journal = "Riv. Nuovo Cim.",
    volume = "30",
    number = "9",
    pages = "387--448",
    year = "2007"
}

@article{Kumericki:2016ehc,
    author = "Kumericki, Kresimir and Liuti, Simonetta and Moutarde, Herve",
    title = "{GPD phenomenology and DVCS fitting}: {Entering the high-precision era}",
    eprint = "1602.02763",
    archivePrefix = "arXiv",
    primaryClass = "hep-ph",
    doi = "10.1140/epja/i2016-16157-3",
    journal = "Eur. Phys. J. A",
    volume = "52",
    number = "6",
    pages = "157",
    year = "2016"
}

@article{Kim:2025mol,
    author = "Kim, June-Young",
    title = "{Chiral-odd generalized parton distributions in the large-Nc limit of QCD: Next-to-leading-order contributions}",
    eprint = "2506.21013",
    archivePrefix = "arXiv",
    primaryClass = "hep-ph",
    reportNumber = "JLAB-THY-25-4386",
    doi = "10.1103/qpzb-1skm",
    journal = "Phys. Rev. D",
    volume = "112",
    number = "7",
    pages = "074032",
    year = "2025"
}

@article{Kim:2024ibz,
    author = "Kim, June-Young and Weiss, Christian",
    title = "{Chiral-odd generalized parton distributions in the large-Nc limit of QCD: Spin-flavor structure, polynomiality, and sum rules}",
    eprint = "2411.17634",
    archivePrefix = "arXiv",
    primaryClass = "hep-ph",
    reportNumber = "JLAB-THY-24-4239",
    doi = "10.1103/PhysRevD.111.074007",
    journal = "Phys. Rev. D",
    volume = "111",
    number = "7",
    pages = "074007",
    year = "2025"
}

@article{Kim:1995bq,
    author = "Kim, Hyun-Chul and Polyakov, Maxim V. and Goeke, Klaus",
    title = "{Nucleon tensor charges in the SU(2) chiral quark - soliton model}",
    eprint = "hep-ph/9509283",
    archivePrefix = "arXiv",
    reportNumber = "RUB-TPII-26-95",
    doi = "10.1103/PhysRevD.53.R4715",
    journal = "Phys. Rev. D",
    volume = "53",
    pages = "4715--4718",
    year = "1996"
}

@article{Ledwig:2010zq,
    author = "Ledwig, Tim and Silva, Antonio and Kim, Hyun-Chul",
    title = "{Anomalous tensor magnetic moments and form factors of the proton in the self-consistent chiral quark-soliton model}",
    eprint = "1007.1355",
    archivePrefix = "arXiv",
    primaryClass = "hep-ph",
    reportNumber = "INHA-NTG-06-2010",
    doi = "10.1103/PhysRevD.82.054014",
    journal = "Phys. Rev. D",
    volume = "82",
    pages = "054014",
    year = "2010"
}

@article{Ledwig:2010tu,
    author = "Ledwig, Tim and Silva, Antonio and Kim, Hyun-Chul",
    title = "{Tensor charges and form factors of SU(3) baryons in the self-consistent SU(3) chiral quark-soliton model}",
    eprint = "1004.3612",
    archivePrefix = "arXiv",
    primaryClass = "hep-ph",
    reportNumber = "INHA-NTG-05-2010",
    doi = "10.1103/PhysRevD.82.034022",
    journal = "Phys. Rev. D",
    volume = "82",
    pages = "034022",
    year = "2010"
}

@article{Ledwig:2011qw,
    author = "Ledwig, Tim and Kim, Hyun-Chul",
    title = "{Transverse strange quark spin structure of the nucleon}",
    eprint = "1107.4952",
    archivePrefix = "arXiv",
    primaryClass = "hep-ph",
    reportNumber = "INHA-NTG-09-2011",
    doi = "10.1103/PhysRevD.85.034041",
    journal = "Phys. Rev. D",
    volume = "85",
    pages = "034041",
    year = "2012"
}

@article{Kim:1996vk,
    author = "Kim, Hyun-Chul and Polyakov, Maxim V. and Goeke, Klaus",
    title = "{Tensor charges of the nucleon in the SU(3) chiral quark soliton model}",
    eprint = "hep-ph/9604442",
    archivePrefix = "arXiv",
    reportNumber = "RUB-TPII-01-96",
    doi = "10.1016/0370-2693(96)01066-0",
    journal = "Phys. Lett. B",
    volume = "387",
    pages = "577--581",
    year = "1996"
}

@article{Belle:2008fdv,
    author = "Seidl, R. and others",
    collaboration = "Belle",
    title = "{Measurement of Azimuthal Asymmetries in Inclusive Production of Hadron Pairs in e+e- Annihilation at s**(1/2) = 10.58-GeV}",
    eprint = "0805.2975",
    archivePrefix = "arXiv",
    primaryClass = "hep-ex",
    reportNumber = "BELLE-2008-14, KEK-2008-8",
    doi = "10.1103/PhysRevD.78.032011",
    journal = "Phys. Rev. D",
    volume = "78",
    pages = "032011",
    year = "2008",
    note = "[Erratum: Phys.Rev.D 86, 039905 (2012)]"
}

@article{Efremov:2004qs,
    author = "Efremov, A. V. and Goeke, K. and Schweitzer, P.",
    title = "{Transversity distribution function in hard scattering of polarized protons and antiprotons in the PAX experiment}", 
    eprint = "hep-ph/0403124",
    archivePrefix = "arXiv",
    doi = "10.1140/epjc/s2004-01854-9",
    journal = "Eur. Phys. J. C",
    volume = "35",
    pages = "207--210",
    year = "2004"
}

@article{PAX:2005leu,
    author = "Barone, Vincenzo and others",
    collaboration = "PAX",
    title = "{Antiproton-proton scattering experiments with polarization}",
    journal ={},
    volume = {},
    pages = {},
    eprint = "hep-ex/0505054",  
    archivePrefix = "arXiv",
    month = "5",
    year = "2005"
}

@article{Pasquini:2006iv,
    author = "Pasquini, B. and Pincetti, M. and Boffi, S.",
    title = "{Drell-Yan processes, transversity and light-cone wavefunctions}",
    eprint = "hep-ph/0612094",
    archivePrefix = "arXiv",
    doi = "10.1103/PhysRevD.76.034020",
    journal = "Phys. Rev. D",
    volume = "76",
    pages = "034020",
    year = "2007"
}

@article{Anselmino:2004ki,
    author = "Anselmino, M. and Barone, V. and Drago, A. and Nikolaev, N. N.",
    title = "{Accessing transversity via J / psi production in polarized p vector anti-p vector interactions}",
    eprint = "hep-ph/0403114",
    archivePrefix = "arXiv",
    doi = "10.1016/j.physletb.2004.05.029",
    journal = "Phys. Lett. B",
    volume = "594",
    pages = "97--104",
    year = "2004"
}

@article{Kahana:1984be,
    author = "Kahana, S. and Ripka, G.",
    title = "{Baryon Density of Quarks Coupled to a Chiral Field}",
    doi = "10.1016/0375-9474(84)90692-4",
    journal = "Nucl. Phys. A",
    volume = "429",
    pages = "462--476",
    year = "1984"
}

@article{dHose:2016mda,
    author = "d'Hose, Nicole and Niccolai, Silvia and Rostomyan, Armine",
    title = "{Experimental overview of Deeply Virtual Compton Scattering}",
    doi = "10.1140/epja/i2016-16151-9",
    journal = "Eur. Phys. J. A",
    volume = "52",
    number = "6",
    pages = "151",
    year = "2016"
}

@article{Pasquini:2005dk,
    author = "Pasquini, B. and Pincetti, M. and Boffi, S.",
    title = "{Chiral-odd generalized parton distributions in constituent quark models}",
    eprint = "hep-ph/0510376",
    archivePrefix = "arXiv",
    doi = "10.1103/PhysRevD.72.094029",
    journal = "Phys. Rev. D",
    volume = "72",
    pages = "094029",
    year = "2005"
}

@article{Burkardt:2005hp,
    author = "Burkardt, Matthias",
    title = "{Transverse deformation of parton distributions and transversity decomposition of angular momentum}",
    eprint = "hep-ph/0505189",
    archivePrefix = "arXiv",
    doi = "10.1103/PhysRevD.72.094020",
    journal = "Phys. Rev. D",
    volume = "72",
    pages = "094020",
    year = "2005"
}

@article{Lorce:2007fa,
    author = "Lorce, Cedric",
    title = "{Tensor charges of light baryons in the Infinite Momentum Frame}",
    eprint = "0708.4168",
    archivePrefix = "arXiv",
    primaryClass = "hep-ph",
    doi = "10.1103/PhysRevD.79.074027",
    journal = "Phys. Rev. D",
    volume = "79",
    pages = "074027",
    year = "2009"
}

@article{Lorce:2011dv,
    author = "Lorce, Cedric and Pasquini, Barbara and Vanderhaeghen, Marc",
    title = "{Unified framework for generalized and transverse-momentum dependent parton distributions within a 3Q light-cone picture of the nucleon}",
    eprint = "1102.4704",
    archivePrefix = "arXiv",
    primaryClass = "hep-ph",
    doi = "10.1007/JHEP05(2011)041",
    journal = "JHEP",
    volume = "05",
    pages = "041",
    year = "2011"
}

@article{Tezgin:2024tfh,
    author = "Tezgin, Kemal and Maynard, Brean and Schweitzer, Peter",
    title = "{Chiral-odd GPDs in the bag model}",
    eprint = "2404.11563",
    archivePrefix = "arXiv",
    primaryClass = "hep-ph",
    doi = "10.1103/PhysRevD.110.054028",
    journal = "Phys. Rev. D",
    volume = "110",
    number = "5",
    pages = "054028",
    year = "2024"
}

@article{Diehl:2005jf,
    author = "Diehl, M. and H{\"a}gler, Ph.",
    title = "{Spin densities in the transverse plane and generalized transversity distributions}",
    eprint = "hep-ph/0504175",
    archivePrefix = "arXiv",
    reportNumber = "DESY-05-061",
    doi = "10.1140/epjc/s2005-02342-6",
    journal = "Eur. Phys. J. C",
    volume = "44",
    pages = "87--101",
    year = "2005"
}

@article{Anselmino:2007fs,
    author = "Anselmino, M. and Boglione, M. and D'Alesio, U. and Kotzinian, A. and Murgia, F. and Prokudin, A. and Turk, C.",
    title = "{Transversity and Collins functions from SIDIS and e+ e- data}",
    eprint = "hep-ph/0701006",
    archivePrefix = "arXiv",
    doi = "10.1103/PhysRevD.75.054032",
    journal = "Phys. Rev. D",
    volume = "75",
    pages = "054032",
    year = "2007"
}

@article{Anselmino:2008jk,
    author = "Anselmino, M. and Boglione, M. and D'Alesio, U. and Kotzinian, A. and Murgia, F. and Prokudin, A. and Melis, S.",
    editor = "Grindhammer, Guenter and Kniehl, Bernd A. and Kramer, Gustav and Ochs, Wolfgang",
    title = "{Update on transversity and Collins functions from SIDIS and e+ e- data}",
    eprint = "0812.4366",
    archivePrefix = "arXiv",
    primaryClass = "hep-ph",
    doi = "10.1016/j.nuclphysbps.2009.03.117",
    journal = "Nucl. Phys. B Proc. Suppl.",
    volume = "191",
    pages = "98--107",
    year = "2009"
}

@article{Anselmino:2013vqa,
    author = "Anselmino, M. and Boglione, M. and D'Alesio, U. and Melis, S. and Murgia, F. and Prokudin, A.",
    title = "{Simultaneous extraction of transversity and Collins functions from new SIDIS and e+e- data}",
    eprint = "1303.3822",
    archivePrefix = "arXiv",
    primaryClass = "hep-ph",
    reportNumber = "JLAB-THY-13-1704",
    doi = "10.1103/PhysRevD.87.094019",
    journal = "Phys. Rev. D",
    volume = "87",
    pages = "094019",
    year = "2013"
}

@article{Kang:2014zza,
    author = "Kang, Zhong-Bo and Prokudin, Alexei and Sun, Peng and Yuan, Feng",
    title = "{Nucleon tensor charge from Collins azimuthal asymmetry measurements}",
    eprint = "1410.4877",
    archivePrefix = "arXiv",
    primaryClass = "hep-ph",
    reportNumber = "JLAB-THY-14-1965",
    doi = "10.1103/PhysRevD.91.071501",
    journal = "Phys. Rev. D",
    volume = "91",
    pages = "071501",
    year = "2015"
}

@article{Ye:2016prn,
    author = "Ye, Zhihong and Sato, Nobuo and Allada, Kalyan and Liu, Tianbo and Chen, Jian-Ping and Gao, Haiyan and Kang, Zhong-Bo and Prokudin, Alexei and Sun, Peng and Yuan, Feng",
    title = "{Unveiling the nucleon tensor charge at Jefferson Lab: A study of the SoLID case}",
    eprint = "1609.02449",
    archivePrefix = "arXiv",
    primaryClass = "hep-ph",
    reportNumber = "JLAB-THY-16-2328",
    doi = "10.1016/j.physletb.2017.01.046",
    journal = "Phys. Lett. B",
    volume = "767",
    pages = "91--98",
    year = "2017"
}

@article{Radici:2015mwa,
    author = "Radici, Marco and Courtoy, A. and Bacchetta, Alessandro and Guagnelli, Marco",
    title = "{Improved extraction of valence transversity distributions from inclusive dihadron production}", 
    eprint = "1503.03495",
    archivePrefix = "arXiv",
    primaryClass = "hep-ph",
    doi = "10.1007/JHEP05(2015)123",
    journal = "JHEP",
    volume = "05",
    pages = "123",
    year = "2015"
}

@article{Lin:2017stx,
    author = "Lin, Huey-Wen and Melnitchouk, W. and Prokudin, Alexei and Sato, N. and Shows, H.",
    title = "{First Monte Carlo Global Analysis of Nucleon Transversity with Lattice QCD Constraints}",
    eprint = "1710.09858",
    archivePrefix = "arXiv",
    primaryClass = "hep-ph",
    reportNumber = "MSUHEP-17-016, JLAB-THY-17-2574",
    doi = "10.1103/PhysRevLett.120.152502",
    journal = "Phys. Rev. Lett.",
    volume = "120",
    number = "15",
    pages = "152502",
    year = "2018"
}

@article{Gao:2025evv,
    author = "Gao, Mei-Sen and Kang, Zhong-Bo and Li, Wanchen and Shao, Ding Yu",
    title = "{Accessing nucleon transversity with one-point energy correlators}",
    journal = "",
    volume = "",
    number = "",
    pages = "",
    eprint = "2509.15809",
    archivePrefix = "arXiv",
    primaryClass = "hep-ph",
    month = "9",
    year = "2025"
}

@article{Cocuzza:2023oam,
    author = "Cocuzza, C. and Metz, A. and Pitonyak, D. and Prokudin, A. and Sato, N. and Seidl, R.",
    collaboration = "JAM",
    title = "{Transversity Distributions and Tensor Charges of the Nucleon: Extraction from Dihadron Production and Their Universal Nature}",
    eprint = "2306.12998",
    archivePrefix = "arXiv",
    primaryClass = "hep-ph",
    reportNumber = "JLAB-THY-23-3859",
    doi = "10.1103/PhysRevLett.132.091901",
    journal = "Phys. Rev. Lett.",
    volume = "132",
    number = "9",
    pages = "091901",
    year = "2024"
}

@article{Cocuzza:2023vqs,
    author = "Cocuzza, C. and Metz, A. and Pitonyak, D. and Prokudin, A. and Sato, N. and Seidl, R.",
    collaboration = "Jefferson Lab Angular Momentum (JAM)",
    title = "{First simultaneous global QCD analysis of dihadron fragmentation functions and transversity parton distribution functions}",
    eprint = "2308.14857",
    archivePrefix = "arXiv",
    primaryClass = "hep-ph",
    reportNumber = "JLAB-THY-23-3901",
    doi = "10.1103/PhysRevD.109.034024",
    journal = "Phys. Rev. D",
    volume = "109",
    number = "3",
    pages = "034024",
    year = "2024"
}

@article{Ralston:1979ys,
    author = "Ralston, John P. and Soper, Davison E.",
    title = "{Production of Dimuons from High-Energy Polarized Proton Proton Collisions}",
    reportNumber = "OITS-100",
    doi = "10.1016/0550-3213(79)90082-8",
    journal = "Nucl. Phys. B",
    volume = "152",
    pages = "109",
    year = "1979"
}

@article{Jaffe:1991ra,
    author = "Jaffe, R. L. and Ji, Xiang-Dong",
    title = "{Chiral odd parton distributions and Drell-Yan processes}",
    reportNumber = "MIT-CTP-2005",
    doi = "10.1016/0550-3213(92)90110-W",
    journal = "Nucl. Phys. B",
    volume = "375",
    pages = "527--560",
    year = "1992"
}

@article{Collins:1992kk,
    author = "Collins, John C.",
    title = "{Fragmentation of transversely polarized quarks probed in transverse momentum distributions}",
    eprint = "hep-ph/9208213",
    archivePrefix = "arXiv",
    reportNumber = "PSU-TH-102",
    doi = "10.1016/0550-3213(93)90262-N",
    journal = "Nucl. Phys. B",
    volume = "396",
    pages = "161--182",
    year = "1993"
}

@article{Aoki:1996pi,
    author = "Aoki, S. and Doui, M. and Hatsuda, T. and Kuramashi, Y.",
    title = "{Tensor charge of the nucleon in lattice QCD}",
    eprint = "hep-lat/9608115",
    archivePrefix = "arXiv",
    reportNumber = "UTHEP-339",
    doi = "10.1103/PhysRevD.56.433",
    journal = "Phys. Rev. D",
    volume = "56",
    pages = "433--436",
    year = "1997"
}

@article{Gupta:2018lvp,
    author = "Gupta, Rajan and Yoon, Boram and Bhattacharya, Tanmoy and Cirigliano, Vincenzo and Jang, Yong-Chull and Lin, Huey-Wen",
    title = "{Flavor diagonal tensor charges of the nucleon from (2+1+1)-flavor lattice QCD}",
    eprint = "1808.07597",
    archivePrefix = "arXiv",
    primaryClass = "hep-lat",
    reportNumber = "LA-UR-18-28007",
    doi = "10.1103/PhysRevD.98.091501",
    journal = "Phys. Rev. D",
    volume = "98",
    number = "9",
    pages = "091501",
    year = "2018"
}

@article{Alexandrou:2017qyt,
    author = "Alexandrou, C. and others",
    title = "{Nucleon scalar and tensor charges using lattice QCD simulations at the physical value of the pion mass}",
    eprint = "1703.08788",
    archivePrefix = "arXiv",
    primaryClass = "hep-lat",
    doi = "10.1103/PhysRevD.95.114514",
    journal = "Phys. Rev. D",
    volume = "95",
    number = "11",
    pages = "114514",
    year = "2017",
    note = "[Erratum: Phys.Rev.D 96, 099906 (2017)]"
}

@article{Alexandrou:2021oih,
    author = "Alexandrou, Constantia and Constantinou, Martha and Hadjiyiannakou, Kyriakos and Jansen, Karl and Manigrasso, Floriano",
    title = "{Flavor decomposition of the nucleon unpolarized, helicity, and transversity parton distribution functions from lattice QCD simulations}",
    eprint = "2106.16065",
    archivePrefix = "arXiv",
    primaryClass = "hep-lat",
    doi = "10.1103/PhysRevD.104.054503",
    journal = "Phys. Rev. D",
    volume = "104",
    number = "5",
    pages = "054503",
    year = "2021"
}

@article{Park:2021ypf,
    author = "Park, Sungwoo and Gupta, Rajan and Yoon, Boram and Mondal, Santanu and Bhattacharya, Tanmoy and Jang, Yong-Chull and Jo\'o, B\'alint and Winter, Frank",
    collaboration = "Nucleon Matrix Elements (NME)",
    title = "{Precision nucleon charges and form factors using (2+1)-flavor lattice QCD}",
    eprint = "2103.05599",
    archivePrefix = "arXiv",
    primaryClass = "hep-lat",
    reportNumber = "LA-UR-21-20526, JLAB-THY-22-3583",
    doi = "10.1103/PhysRevD.105.054505",
    journal = "Phys. Rev. D",
    volume = "105",
    number = "5",
    pages = "054505",
    year = "2022"
}

@article{QCDSF:2006tkx,
    author = {G{\"o}ckeler, M. and H{\"a}gler, Ph. and Horsley, R. and Nakamura, Y. and Pleiter, D. and Rakow, P. E. L. and Sch{\"a}fer, A. and Schierholz, G. and St{\"u}ben, H. and Zanotti, J. M.},
    collaboration = "QCDSF, UKQCD",
    title = "{Transverse spin structure of the nucleon from lattice QCD simulations}",
    eprint = "hep-lat/0612032",
    archivePrefix = "arXiv",
    reportNumber = "DESY-06-245, EDINBURGH-2006-41, TUM-T39-06-16",
    doi = "10.1103/PhysRevLett.98.222001",
    journal = "Phys. Rev. Lett.",
    volume = "98",
    pages = "222001",
    year = "2007"
}

@article{Ghim:2025gqo,
    author = "Ghim, Nam-Yong and Won, Ho-Yeon and Kim, June-Young and Kim, Hyun-Chul",
    title = "{Nucleon tensor form factors at large Nc}",
    eprint = "2501.12241",
    archivePrefix = "arXiv",
    primaryClass = "hep-ph",
    reportNumber = "INHA-NTG-02/2025, JLAB-THY-25-4080",
    doi = "10.1103/PhysRevD.111.074024",
    url = "https://link.aps.org/doi/10.1103/PhysRevD.111.074024",
    journal = "Phys. Rev. D",
    volume = "111",
    number = "7",
    pages = "074024",
    year = "2025"
}

@article{Schafer:1996wv,
    author = {Sch\"afer, Thomas and Shuryak, Edward V.},
    title = "{Instantons in QCD}",
    eprint = "hep-ph/9610451",
    archivePrefix = "arXiv",
    reportNumber = "DOE-ER-40561-293, INT-96-00-150",
    doi = "10.1103/RevModPhys.70.323",
    journal = "Rev. Mod. Phys.",
    volume = "70",
    pages = "323--426",
    year = "1998"
}

@article{Diakonov:2002fq,
    author = "Diakonov, Dmitri",
    title = "{Instantons at work}",
    eprint = "hep-ph/0212026",
    archivePrefix = "arXiv",
    reportNumber = "NORDITA-2002-74-HE",
    doi = "10.1016/S0146-6410(03)90014-7",
    journal = "Prog. Part. Nucl. Phys.",
    volume = "51",
    pages = "173--222",
    year = "2003"
}

@article{Diakonov:1987ty,
    author = "Diakonov, Dmitri and Petrov, V. Yu. and Pobylitsa, P. V.",
    title = "{A Chiral Theory of Nucleons}",
    reportNumber = "LENINGRAD-87-1297",
    doi = "10.1016/0550-3213(88)90443-9",
    journal = "Nucl. Phys. B",
    volume = "306",
    pages = "809",
    year = "1988"
}

@article{Witten:1979kh,
    author = "Witten, Edward",
    title = "{Baryons in the $1/N$ Expansion}",
    reportNumber = "HUTP-79-A007",
    doi = "10.1016/0550-3213(79)90232-3",
    journal = "Nucl. Phys. B",
    volume = "160",
    pages = "57--115",
    year = "1979"
}

@article{Witten:1983tw,
    author = "Witten, Edward",
    title = "{Global Aspects of Current Algebra}",
    reportNumber = "PRINT-83-0262 (PRINCETON)",
    doi = "10.1016/0550-3213(83)90063-9",
    journal = "Nucl. Phys. B",
    volume = "223",
    pages = "422--432",
    year = "1983"
}

@article{Christov:1995vm,
    author = "Christov, Chr. V. and Blotz, A. and Kim, Hyun-Chul and Pobylitsa, P. and Watabe, T. and Meissner, T. and Ruiz Arriola, E. and Goeke, K.",
    title = "{Baryons as nontopological chiral solitons}",
    eprint = "hep-ph/9604441",
    archivePrefix = "arXiv",
    reportNumber = "RUB-TPII-32-95",
    doi = "10.1016/0146-6410(96)00057-9",
    journal = "Prog. Part. Nucl. Phys.",
    volume = "37",
    pages = "91--191",
    year = "1996"
}

@article{Wakamatsu:2008ki,
    author = "Wakamatsu, M.",
    title = "{Chiral-odd GPDs, transversity decomposition of angular momentum, and tensor charges of the nucleon}",
    eprint = "0811.4196",
    archivePrefix = "arXiv",
    primaryClass = "hep-ph",
    reportNumber = "OU-HET-619",
    doi = "10.1103/PhysRevD.79.014033",
    journal = "Phys. Rev. D",
    volume = "79",
    pages = "014033",
    year = "2009"
}

@article{WAKAMATSU2007398,
  title = {Comparative analysis of the transversities and the longitudinally polarized distribution functions of the nucleon},
  journal = {Physics Letters B},
  volume = {653},
  number = {5},
  pages = {398-403},
  year = {2007},
  issn = {0370-2693},
  doi = {https://doi.org/10.1016/j.physletb.2007.08.013},
  url = {https://www.sciencedirect.com/science/article/pii/S0370269307009689},
  author = {M. Wakamatsu},
}

@article{He:1994gz,
    author = "He, Han-xin and Ji, Xiang-Dong",
    title = "{The Nucleon's tensor charge}",
    eprint = "hep-ph/9412235",
    archivePrefix = "arXiv",
    reportNumber = "MIT-CTP-2380",
    doi = "10.1103/PhysRevD.52.2960",
    journal = "Phys. Rev. D",
    volume = "52",
    pages = "2960--2963",
    year = "1995"
}

@article{Alexandrou:2019brg,
    author = "Alexandrou, C. and Bacchio, S. and Constantinou, M. and Finkenrath, J. and Hadjiyiannakou, K. and Jansen, K. and Koutsou, G. and Vaquero Aviles-Casco, A.",
    title = "{Nucleon axial, tensor, and scalar charges and $\sigma$-terms in lattice QCD}",
    eprint = "1909.00485",
    archivePrefix = "arXiv",
    primaryClass = "hep-lat",
    doi = "10.1103/PhysRevD.102.054517",
    journal = "Phys. Rev. D",
    volume = "102",
    number = "5",
    pages = "054517",
    month = "Sep.",
    year = "2020"
}

@article{Bhattacharya:2015wna,
  title = {Isovector and isoscalar tensor charges of the nucleon from lattice QCD},
  author = {Bhattacharya, Tanmoy and Cirigliano, Vincenzo and Cohen, Saul D. and Gupta, Rajan and Joseph, Anosh and Lin, Huey-Wen and Yoon, Boram},
  collaboration = {Precision Neutron Decay Matrix Elements (PNDME) Collaboration},
  journal = {Phys. Rev. D},
  volume = {92},
  issue = {9},
  pages = {094511},
  numpages = {21},
  year = {2015},
  month = {Nov},
  publisher = {American Physical Society},
  doi = {10.1103/PhysRevD.92.094511},
  url = {https://link.aps.org/doi/10.1103/PhysRevD.92.094511}
  }

@article{Alexandrou:2022dtc,
  title = {Moments of the nucleon transverse quark spin densities using lattice QCD},
  author = {Alexandrou, C. and Bacchio, S. and Constantinou, M. and Dimopoulos, P. and Finkenrath, J. and Frezzotti, R. and Hadjiyiannakou, K. and Jansen, K. and Kostrzewa, B. and Koutsou, G. and Spanoudes, G. and Urbach, C.},
  journal = {Phys. Rev. D},
  volume = {107},
  issue = {5},
  pages = {054504},
  numpages = {15},
  year = {2023},
  month = {Mar},
  publisher = {American Physical Society},
  doi = {10.1103/PhysRevD.107.054504},
  url = {https://link.aps.org/doi/10.1103/PhysRevD.107.054504}
}

@article{Erkol:2011iw,
    author = "Erkol, Guray and Ozpineci, Altug",
    title = "{Tensor form factors of nucleon in QCD}",
    eprint = "1107.4584",
    archivePrefix = "arXiv",
    primaryClass = "hep-ph",
    doi = "10.1016/j.physletb.2011.09.089",
    journal = "Phys. Lett. B",
    volume = "704",
    pages = "551--558",
    year = "2011"
}

@article{Bukhvostov:1984rns,
    author = "Bukhvostov, A. P. and Kuraev, E. A. and Lipatov, L. N.",
    title = "{Deep inelastic scattering by a polarized target in quantum chromodynamics}",
    journal = "Sov. Phys. JETP",
    volume = "60",
    number = "1",
    pages = "22--32",
    year = "1984"
}

@article{Artru:1989zv,
    author = "Artru, Xavier and Mekhfi, Mustapha",
    title = "{Transversely Polarized Parton Densities, their Evolution and their Measurement}",
    reportNumber = "LPTHE-ORSAY-89-25",
    doi = "10.1007/BF01556280",
    journal = "Z. Phys. C",
    volume = "45",
    pages = "669",
    year = "1990"
}

@article{Diehl:2023nmm,
    author = "Diehl, Stefan",
    title = "{Experimental exploration of the 3D nucleon structure}",
    doi = "10.1016/j.ppnp.2023.104069",
    journal = "Prog. Part. Nucl. Phys.",
    volume = "133",
    pages = "104069",
    year = "2023"
}

@article{Cortes:1991ja,
    author = "Cortes, J. L. and Pire, B. and Ralston, J. P.",
    title = "{Measuring the transverse polarization of quarks in the proton}",
    reportNumber = "CPTH-A048-0491",
    doi = "10.1007/BF01565099",
    journal = "Z. Phys. C",
    volume = "55",
    pages = "409--416",
    year = "1992"
}

@article{Barone:2010zz,
    author = "Barone, Vincenzo and Bradamante, Franco and Martin, Anna",
    title = "{Transverse-spin and transverse-momentum effects in high-energy processes}",
    eprint = "1011.0909",
    archivePrefix = "arXiv",
    primaryClass = "hep-ph",
    doi = "10.1016/j.ppnp.2010.07.003",
    journal = "Prog. Part. Nucl. Phys.",
    volume = "65",
    pages = "267--333",
    year = "2010"
}

@article{Aidala:2012mv,
    author = "Aidala, Christine A. and Bass, Steven D. and Hasch, Delia and Mallot, Gerhard K.",
    title = "{The Spin Structure of the Nucleon}",
    eprint = "1209.2803",
    archivePrefix = "arXiv",
    primaryClass = "hep-ph",
    doi = "10.1103/RevModPhys.85.655",
    journal = "Rev. Mod. Phys.",
    volume = "85",
    pages = "655--691",
    year = "2013"
}

@article{Rodekamp:2023wpe,
    author = "Rodekamp, Marcel and Engelhardt, Michael and Green, Jeremy R. and Krieg, Stefan and Liuti, Simonetta and Meinel, Stefan and Negele, John W. and Pochinsky, Andrew and Syritsyn, Sergey",
    title = "{Moments of nucleon unpolarized, polarized, and transversity parton distribution functions from lattice QCD at the physical point}",
    eprint = "2401.05360",
    archivePrefix = "arXiv",
    primaryClass = "hep-lat",
    reportNumber = "DESY-23-212",
    doi = "10.1103/PhysRevD.109.074508",
    journal = "Phys. Rev. D",
    volume = "109",
    number = "7",
    pages = "074508",
    year = "2024"
}

@article{Yamanaka:2018uud,
    author = "Yamanaka, Nodoka and Hashimoto, Shoji and Kaneko, Takashi and Ohki, Hiroshi",
    collaboration = "JLQCD",
    title = "{Nucleon charges with dynamical overlap fermions}",
    eprint = "1805.10507",
    archivePrefix = "arXiv",
    primaryClass = "hep-lat",
    reportNumber = "KEK-CP-365",
    doi = "10.1103/PhysRevD.98.054516",
    journal = "Phys. Rev. D",
    volume = "98",
    number = "5",
    pages = "054516",
    year = "2018"
}

@article{Pospelov:2005pr,
    author = "Pospelov, Maxim and Ritz, Adam",
    title = "{Electric dipole moments as probes of new physics}",
    eprint = "hep-ph/0504231",
    archivePrefix = "arXiv",
    doi = "10.1016/j.aop.2005.04.002",
    journal = "Annals Phys.",
    volume = "318",
    pages = "119--169",
    year = "2005"
}

@article{Gonzalez-Alonso:2018omy,
    author = "Gonz{\'a}lez-Alonso, Martin and Naviliat-Cuncic, Oscar and Severijns, Nathal",
    title = "{New physics searches in nuclear and neutron $\beta$ decay}",
    eprint = "1803.08732",
    archivePrefix = "arXiv",
    primaryClass = "hep-ph",
    reportNumber = "CERN-TH-2018-050",
    doi = "10.1016/j.ppnp.2018.08.002",
    journal = "Prog. Part. Nucl. Phys.",
    volume = "104",
    pages = "165--223",
    year = "2019"
}

@article{Erler:2004cx,
    author = "Erler, Jens and Ramsey-Musolf, Michael J.",
    title = "{Low energy tests of the weak interaction}",
    eprint = "hep-ph/0404291",
    archivePrefix = "arXiv",
    reportNumber = "FT-2004-02",
    doi = "10.1016/j.ppnp.2004.08.001",
    journal = "Prog. Part. Nucl. Phys.",
    volume = "54",
    pages = "351--442",
    year = "2005"
}

@article{Severijns:2006dr,
    author = "Severijns, Nathal and Beck, Marcus and Naviliat-Cuncic, Oscar",
    title = "{Tests of the standard electroweak model in beta decay}",
    eprint = "nucl-ex/0605029",
    archivePrefix = "arXiv",
    doi = "10.1103/RevModPhys.78.991",
    journal = "Rev. Mod. Phys.",
    volume = "78",
    pages = "991--1040",
    year = "2006"
}

@article{Cirigliano:2013xha,
    author = "Cirigliano, Vincenzo and Gardner, Susan and Holstein, Barry",
    title = "{Beta Decays and Non-Standard Interactions in the LHC Era}",
    eprint = "1303.6953",
    archivePrefix = "arXiv",
    primaryClass = "hep-ph",
    doi = "10.1016/j.ppnp.2013.03.005",
    journal = "Prog. Part. Nucl. Phys.",
    volume = "71",
    pages = "93--118",
    year = "2013"
}

@article{Courtoy:2015haa,
    author = "Courtoy, Aurore and Bae{\ss}ler, Stefan and Gonz{\'a}lez-Alonso, Mart{\'\i}n and Liuti, Simonetta",
    title = "{Beyond-Standard-Model Tensor Interaction and Hadron Phenomenology}",
    eprint = "1503.06814",
    archivePrefix = "arXiv",
    primaryClass = "hep-ph",
    doi = "10.1103/PhysRevLett.115.162001",
    journal = "Phys. Rev. Lett.",
    volume = "115",
    pages = "162001",
    year = "2015"
}

@article{Bishara:2017pfq,
    author = "Bishara, Fady and Brod, Joachim and Grinstein, Benjamin and Zupan, Jure",
    title = "{From quarks to nucleons in dark matter direct detection}",
    eprint = "1707.06998",
    archivePrefix = "arXiv",
    primaryClass = "hep-ph",
    reportNumber = "DO-TH-17-10, OUTP-17-07P, CERN-TH-2017-157",
    doi = "10.1007/JHEP11(2017)059",
    journal = "JHEP",
    volume = "11",
    pages = "059",
    year = "2017"
}

@article{Wang:2025nsd,
    author = "Wang, Ji-Hao and Hu, Zhi-Cheng and Ji, Xiangdong and Jiang, Xiangyu and Su, Yushan and Sun, Peng and Yang, Yi-Bo",
    title = "{Accurate nucleon iso-vector scalar and tensor charge at physical point}",
    journal = "",
    volume = "",
    pages = "",
    eprint = "2511.02326",
    archivePrefix = "arXiv",
    primaryClass = "hep-lat",
    month = "11",
    year = "2025"
}

@article{Ma:2024aoc,
    author = "Ma, Kai",
    title = "{Exploring four fermion contact couplings of a dark fermion and an electron at hadron colliders and direct detection experiments}",
    eprint = "2404.06419",
    archivePrefix = "arXiv",
    primaryClass = "hep-ph",
    doi = "10.1016/j.dark.2025.102090",
    journal = "Phys. Dark Univ.",
    volume = "50",
    pages = "102090",
    year = "2025"
}

@article{Antel:2023hkf,
    author = "Antel, C. and others",
    title = "{Feebly-interacting particles: FIPs 2022 Workshop Report}",
    eprint = "2305.01715",
    archivePrefix = "arXiv",
    primaryClass = "hep-ph",
    reportNumber = "CERN-TH-2023-061, DESY-23-050, FERMILAB-PUB-23-149-PPD, INFN-23-14-LNF, JLAB-PHY-23-3789, LA-UR-23-21432, MITP-23-015",
    doi = "10.1140/epjc/s10052-023-12168-5",
    journal = "Eur. Phys. J. C",
    volume = "83",
    number = "12",
    pages = "1122",
    year = "2023"
}

@article{Goodman:2010ku,
    author = "Goodman, Jessica and Ibe, Masahiro and Rajaraman, Arvind and Shepherd, William and Tait, Tim M. P. and Yu, Hai-Bo",
    title = "{Constraints on Dark Matter from Colliders}",
    eprint = "1008.1783",
    archivePrefix = "arXiv",
    primaryClass = "hep-ph",
    reportNumber = "UCI-HEP-TR-2010-15",
    doi = "10.1103/PhysRevD.82.116010",
    journal = "Phys. Rev. D",
    volume = "82",
    pages = "116010",
    year = "2010"
}

@article{Brod:2017bsw,
    author = "Brod, Joachim and Gootjes-Dreesbach, Aaron and Tammaro, Michele and Zupan, Jure",
    title = "{Effective Field Theory for Dark Matter Direct Detection up to Dimension Seven}",
    eprint = "1710.10218",
    archivePrefix = "arXiv",
    primaryClass = "hep-ph",
    reportNumber = "DO-TH-17-21, DO-TH 17/21",
    doi = "10.1007/JHEP10(2018)065",
    journal = "JHEP",
    volume = "10",
    pages = "065",
    year = "2018",
    note = "[Erratum: JHEP 07, 012 (2023)]"
}

@article{Adler:1975he,
    author = "Adler, Stephen L. and Colglazier, Jr., E. W. and Healy, J. B. and Karliner, Inga and Lieberman, Judy and Ng, Yee Jack and Tsao, Hung-Sheng",
    title = "{Renormalization Constants for Scalar, Pseudoscalar, and Tensor Currents}",
    reportNumber = "COO-2220-38",
    doi = "10.1103/PhysRevD.11.3309",
    journal = "Phys. Rev. D",
    volume = "11",
    pages = "3309",
    year = "1975"
}

@article{Liang:2025kkl,
    author = "Liang, Jin-Han and Liao, Yi and Ma, Xiao-Dong and Wang, Hao-Lin",
    title = "{Systematic investigation on vector dark matter-nucleus scattering in effective field theories}",
    eprint = "2501.13501",
    archivePrefix = "arXiv",
    primaryClass = "hep-ph",
    doi = "10.1103/PhysRevD.111.095033",
    journal = "Phys. Rev. D",
    volume = "111",
    number = "9",
    pages = "095033",
    year = "2025"
}

@article{Glick-Magid:2023uhk,
    author = "Glick-Magid, Ayala",
    title = "{Nonrelativistic nuclear reduction for tensor couplings in dark matter direct detection and {\ensuremath{\mu}}{\textrightarrow}e conversion}",
    eprint = "2312.08339",
    archivePrefix = "arXiv",
    primaryClass = "hep-ph",
    reportNumber = "INT-PUB-24-028, NT@UW-24-13",
    doi = "10.1103/PhysRevD.110.L051701",
    journal = "Phys. Rev. D",
    volume = "110",
    number = "5",
    pages = "L051701",
    year = "2024"
}

@article{Kaur:2025ssu,
    author = "Kaur, Navpreet and Dahiya, Harleen",
    title = "{Chiral-odd generalized parton distributions for the low-lying octet baryons}",
    eprint = "2506.02464",
    archivePrefix = "arXiv",
    primaryClass = "hep-ph",
    doi = "10.1103/rt3y-2b8g",
    journal = "Phys. Rev. D",
    volume = "112",
    number = "7",
    pages = "074024",
    year = "2025"
}

@article{Kaur:2023lun,
    author = "Kaur, Satvir and Xu, Siqi and Mondal, Chandan and Zhao, Xingbo and Vary, James P.",
    collaboration = "BLFQ",
    title = "{Spatial imaging of proton via leading-twist nonskewed GPDs with basis light-front quantization}",
    eprint = "2307.09869",
    archivePrefix = "arXiv",
    primaryClass = "hep-ph",
    doi = "10.1103/PhysRevD.109.014015",
    journal = "Phys. Rev. D",
    volume = "109",
    number = "1",
    pages = "014015",
    year = "2024"
}

@article{Luan:2024vgv,
    author = "Luan, Xiaoyan and Lu, Zhun",
    title = "{Chiral-odd generalized parton distributions of sea quarks at {\ensuremath{\xi}}=0 in the light-cone quark model}",
    eprint = "2404.13962",
    archivePrefix = "arXiv",
    primaryClass = "hep-ph",
    doi = "10.1103/PhysRevD.110.034021",
    journal = "Phys. Rev. D",
    volume = "110",
    number = "3",
    pages = "034021",
    year = "2024"
}

@article{Aliev:2011ku,
    author = "Aliev, T. M. and Azizi, K. and Savci, M.",
    title = "{Nucleon tensor form factors induced by isovector and isoscalar currents in QCD}",
    eprint = "1108.2019",
    archivePrefix = "arXiv",
    primaryClass = "hep-ph",
    doi = "10.1103/PhysRevD.84.076005",
    journal = "Phys. Rev. D",
    volume = "84",
    pages = "076005",
    year = "2011"
}

@article{Callan:1977gz,
    author = "Callan, Jr., Curtis G. and Dashen, Roger F. and Gross, David J.",
    editor = "Shifman, Mikhail A.",
    title = "{Toward a Theory of the Strong Interactions}",
    reportNumber = "COO-2220-115",
    doi = "10.1103/PhysRevD.17.2717",
    journal = "Phys. Rev. D",
    volume = "17",
    pages = "2717",
    year = "1978"
}

@article{Carlitz:1978yj,
    author = "Carlitz, Robert D. and Creamer, Dennis B.",
    title = "{Light Quarks and Instantons}",
    reportNumber = "PITT-199",
    doi = "10.1016/0003-4916(79)90133-7",
    journal = "Annals Phys.",
    volume = "118",
    pages = "429",
    year = "1979"
}

@article{Shuryak:1981ff,
    author = "Shuryak, Edward V.",
    title = "{The Role of Instantons in Quantum Chromodynamics. 1. Physical Vacuum}",
    reportNumber = "IYF 81-118",
    doi = "10.1016/0550-3213(82)90478-3",
    journal = "Nucl. Phys. B",
    volume = "203",
    pages = "93",
    year = "1982"
}

@article{Yang:2016qdz,
    author = "Yang, Ghil-Seok and Kim, Hyun-Chul and Polyakov, Maxim V. and Prasza{\l}owicz, Micha{\l}",
    title = "{Pion mean fields and heavy baryons}",
    eprint = "1607.07089",
    archivePrefix = "arXiv",
    primaryClass = "hep-ph",
    reportNumber = "INHA-NTG-01-2016",
    doi = "10.1103/PhysRevD.94.071502",
    journal = "Phys. Rev. D",
    volume = "94",
    pages = "071502",
    year = "2016"
}

@article{Kim:2018cxv,
    author = "Kim, Hyun-Chul",
    title = "{Heavy baryons in a pion mean-field approach: A brief review}",
    eprint = "1804.04393",
    archivePrefix = "arXiv",
    primaryClass = "hep-ph",
    reportNumber = "INHA-NTG-05/2018, INHA-NTG-05-2018",
    doi = "10.3938/jkps.73.165",
    journal = "J. Korean Phys. Soc.",
    volume = "73",
    number = "2",
    pages = "165--178",
    year = "2018"
}

@article{Kim:2019rcx,
    author = "Kim, June-Young and Kim, Hyun-Chul",
    title = "{Improved pion mean fields and masses of singly heavy baryons}",
    eprint = "1909.00123",
    archivePrefix = "arXiv",
    primaryClass = "hep-ph",
    reportNumber = "INHA-NTG-08/2019",
    doi = "10.1093/ptep/ptaa037",
    journal = "PTEP",
    volume = "2020",
    number = "4",
    pages = "043D03",
    year = "2020"
}

@article{Suh:2022atr,
    author = "Suh, Jung-Min and Kim, June-Young and Yang, Ghil-Seok and Kim, Hyun-Chul",
    title = "{Quark spin content of SU(3) light and singly heavy baryons}",
    eprint = "2208.04447",
    archivePrefix = "arXiv",
    primaryClass = "hep-ph",
    reportNumber = "INHA-NTG-09/2022",
    doi = "10.1103/PhysRevD.106.054032",
    journal = "Phys. Rev. D",
    volume = "106",
    number = "5",
    pages = "054032",
    year = "2022"
}

\end{document}